\documentclass[twocolumn]{aastex631} 

\usepackage{newtxtext,newtxmath,comment} 
\usepackage[normalem]{ulem}

\usepackage[T1]{fontenc}
\usepackage{ae,aecompl}
\usepackage{xcolor}
\usepackage{graphicx}
\usepackage{amsmath}
\usepackage{enumitem}
\usepackage{url}
\usepackage{tasks}
\usepackage[utf8]{inputenc}
\usepackage{CJKutf8}
\usepackage{makecell}

\def\mdot{\mathrm{M}_\odot}
\def\mtot{M_{\rm TOT}}
\def\mstar{M_*}
\def\ew{$\widebar{e}_w$}
\def\kw{$\kappa_w$}
\def\om{$\Omega_{\rm m}$}
\def\s8{$\sigma_8$}
\def\agn{$\epsilon_{f,\mathrm{high}}$}
\def \alphaslope{\alpha_{\scriptscriptstyle{\text{1.5~kpc}}}}

\makeatletter
\let\frontmatter@title@above=\relax
\makeatother

\begin{document}

\title[DREAMS CDM Satellites]{The DREAMS Project:\\
Disentangling the Impact of Halo-to-Halo Variance and Baryonic Feedback on \\
Milky Way Satellite Galaxies}

\correspondingauthor{Jonah C. Rose} \\
\email{jr8952@princeton.edu}

 \author[0000-0002-2628-0237]{Jonah C. Rose}
\affiliation{Department of Physics, Princeton University, Princeton, NJ 08544, USA}
\affiliation{Center for Computational Astrophysics, Flatiron Institute, 162 5th Avenue, New York, NY 10010, USA}

\author[0000-0002-8495-8659]{Mariangela Lisanti}
\affiliation{Department of Physics, Princeton University, Princeton, NJ 08544, USA}
\affiliation{Center for Computational Astrophysics, Flatiron Institute, 162 5th Avenue, New York, NY 10010, USA}

\author[0000-0002-5653-0786]{Paul Torrey}
\affiliation{Department of Astronomy, University of Virginia, 530 McCormick Road, Charlottesville, VA 22904}
\affiliation{Virginia Institute for Theoretical Astronomy, University of Virginia, Charlottesville, VA 22904, USA}
\affiliation{The NSF-Simons AI Institute for Cosmic Origins, USA}

\author[0000-0002-4816-0455]{Francisco Villaescusa-Navarro}
\affiliation{Center for Computational Astrophysics, Flatiron Institute, 162 5th Avenue, New York, NY 10010, USA}

\author[0000-0002-8111-9884]{Alex M. Garcia}
\affiliation{Department of Astronomy, University of Virginia, 530 McCormick Road, Charlottesville, VA 22904}
\affiliation{Virginia Institute for Theoretical Astronomy, University of Virginia, Charlottesville, VA 22904, USA}
\affiliation{The NSF-Simons AI Institute for Cosmic Origins, USA}

\author[0000-0003-0777-4618]{Arya~Farahi}
\affiliation{Departments of Statistics and Data Sciences, University of Texas at Austin, Austin, TX 78757, USA}
\affiliation{The NSF-Simons AI Institute for Cosmic Origins, USA}

\author[0000-0001-5522-5029]{Carrie~Filion}
\affiliation{Center for Computational Astrophysics, Flatiron Institute, 162 5th Avenue, New York, NY 10010, USA}

\author[0000-0002-0372-3736]{Alyson M. Brooks}
\affiliation{Department of Physics \& Astronomy, Rutgers, the State University of New Jersey, Piscataway, NJ 08854, USA}

\author[0000-0002-3204-1742]{Nitya Kallivayalil} 
\affiliation{Department of Astronomy, University of Virginia, 530 McCormick Road, Charlottesville, VA 22904}
\affiliation{The NSF-Simons AI Institute for Cosmic Origins, USA}

\author[0000-0003-4004-2451]{Kassidy E. Kollmann}
\affiliation{Department of Physics, Princeton University, Princeton, NJ 08544, USA}

\author[0009-0000-8180-9044]{Ethan Lilie} 
\affiliation{Department of Physics, Princeton University, Princeton, NJ 08544, USA}

\author[0000-0001-9592-4190]{Jiaxuan Li (\begin{CJK}{UTF8}{gbsn}李嘉轩\end{CJK}\!\!)} 
\affiliation{Department of Astrophysical Sciences, 4 Ivy Lane, Princeton University, Princeton, NJ 08540, USA}

\author[0009-0009-0239-8706]{Olivia Mostow}
\affiliation{The Oskar Klein Centre, Department of Physics, Stockholm University, Albanova University Center, 106 91 Stockholm, Sweden}

\author[0000-0001-7831-4892]{Akaxia Cruz}
\affiliation{Center for Computational Astrophysics, Flatiron Institute, 162 5th Avenue, New York, NY 10010, USA}
\affiliation{Department of Physics, Princeton University, Princeton, NJ 08544, USA}

\author[0000-0001-6189-8457]{Tri Nguyen} 
\affiliation{Center for Interdisciplinary Exploration and Research in Astrophysics, Northwestern University, 1800 Sherman Ave, Evanston, IL 60201}
\affiliation{NSF-Simons AI Institute for the Sky, 172 E. Chestnut St., Chicago, IL 60611, USA}

\author[0000-0002-7638-7454]{Sandip Roy} 
\affiliation{Department of Astronomy \& Astrophysics, University of California, San Diego, La Jolla, CA 92093, USA}

\author[0000-0002-6021-8760]{Andrew B. Pace}
\affiliation{Department of Astronomy, University of Virginia, 
530 McCormick Road, 
Charlottesville, VA 22904}

\author[0009-0002-1233-2013]{Niusha Ahvazi} 
\affiliation{Department of Astronomy, University of Virginia, 530 McCormick Road, Charlottesville, VA 22904}
\affiliation{Virginia Institute for Theoretical Astronomy, University of Virginia, Charlottesville, VA 22904, USA}
\affiliation{The NSF-Simons AI Institute for Cosmic Origins, USA}

\author[0000-0002-7968-2088]{Stephanie O'Neil} 
\affiliation{Department of Physics \& Astronomy, University of Pennsylvania, Philadelphia, PA 19104, USA}
\affiliation{Department of Physics, Princeton University, Princeton, NJ 08544, USA}

\author[0000-0002-6196-823X]{Xuejian Shen} 
\affiliation{Department of Physics and Kavli Institute for Astrophysics and Space Research, Massachusetts Institute of Technology, Cambridge, MA 02139, USA}

\author[0000-0002-7939-2988]{Francis-Yan Cyr-Racine}
\affiliation{Department of Physics and Astronomy, University of New Mexico, 210 Yale Blvd NE, Albuquerque, NM 87106, USA}

\author[0000-0003-0872-7098]{Adrian M. Price-Whelan}
\affiliation{Center for Computational Astrophysics, Flatiron Institute, 162 5th Avenue, New York, NY 10010, USA}

\author[0000-0002-7007-9725]{Marla Geha} 
\affiliation{Department of Astronomy, Yale University, New Haven, CT 06520, USA}

\author[0000-0003-2806-1414]{Lina Necib} 
\affiliation{Department of Physics and Kavli Institute for Astrophysics and Space Research, Massachusetts Institute of Technology, Cambridge, MA 02139, USA}
\affiliation{The NSF AI Institute for Artificial Intelligence and Fundamental Interactions, Cambridge, MA 02139, USA}

\author[0000-0001-8593-7692]{Mark Vogelsberger}
\affiliation{Department of Physics $\&$ Kavli Institute for Astrophysics and Space Research, Massachusetts Institute of Technology, Cambridge, MA 02139, USA}

\author[0000-0002-8984-0465]{Julian B.~Mu\~noz}
\affiliation{Department of Astronomy, University of Texas at Austin, Austin, TX 78757, USA}

\author[0000-0002-1264-2006]{Julianne J.~Dalcanton}
\affiliation{Center for Computational Astrophysics, Flatiron Institute, 162 5th Avenue, New York, NY 10010, USA}
\affiliation{Department of Astronomy, University of Washington, Box 351580, Seattle, WA 98195, USA}

\begin{abstract}
We analyze the properties of satellite galaxies around 1,024 Milky Way-mass hosts from the DREAMS Project, simulated within a $\Lambda$CDM cosmology. 
Utilizing the TNG galaxy-formation model, the DREAMS simulations incorporate both baryonic physics and cosmological uncertainties for a large sample of galaxies with diverse environments and formation histories.
We investigate the relative impact of the physical uncertainty from the galaxy-formation model on predicted satellite properties using four metrics: the satellite stellar mass function, radial distribution, inner slope of dark matter density profile, and stellar half-light radius.
We compare these predictions to observations from the SAGA Survey and the DREAMS N-body simulations and find that uncertainties from baryonic physics modeling are subdominant to the scatter arising from halo-to-halo variance.
Where baryonic modeling does affect satellites, the supernova wind energy has the largest effect on the satellite properties that we investigate.
Specifically, increased supernova wind energy suppresses the stellar mass of satellites and results in more extended stellar half-light radii.
The adopted wind speed has only a minor impact, and other astrophysical and cosmological parameters show no measurable effect. 
Our findings highlight the robustness of satellite properties against uncertainties in baryonic physics modeling. 
\end{abstract}


\section{Introduction}
\label{sec:intro}

Cosmological hydrodynamical simulations provide abundant data that have revolutionized our understanding of the physical processes that govern the formation and growth of galaxies~\citep{2020Vogelsberger}.
Modern simulations are able to match a variety of galaxy properties, including the evolving galaxy stellar mass functions~\citep[e.g.,][]{2014Genel, 2018Pillepichb, Anbajagane2020stellar}, size evolution~\citep{Brooks2011, 2016Wellons, 2018Genel}, and metal content of galaxies~\citep{Christensen2018, 2019Torrey, 2024Garcia, 2025Garcia}.
Critically, the increasing resolution of these simulations allows for detailed analysis of the satellite galaxies that orbit hosts like our own Milky Way~(MW)~\citep{2015Sales, 2016Nierenberg, Applebaum2021}.

These simulations are especially useful for studying satellites around the MW and other nearby galaxies, as these systems are particularly sensitive to the physical processes governing galaxy formation---such as feedback, environmental effects, and reionization~\citep[e.g.,][]{2010Okamoto}. Their resolved stellar populations and well-characterized orbits provide stringent tests of theoretical models, especially in regimes where astrophysical processes and/or feedback dominate galaxy evolution.
This provides an opportunity to test model assumptions, such as the impact of star formation prescriptions/implementations~\citep{2024Katz, 2025Kang},  stellar feedback~\citep{2002Benson, 2011Wadepuhl, 2018Smith}, reionization~\citep{2010Busha, 2012Lunnan}, and  the interaction of satellites with their host galaxy environment~\citep[e.g.,][]{Zolotov2012, 2012Wetzel, Brooks2014, 2018Simpson, Akins2021, 2022Font, Samuel2022}.
However, most high-resolution simulations only implement a single instance of these assumptions, with some analyses comparing a handful of simulation codes to understand the uncertainties on the galaxy properties---mostly during model calibration studies~\citep[e.g.,][]{2000Kay, 2014Kim, 2014Torrey, 2015Crain, FIRE2, 2024Katz, 2025Pakmor}.
The general lack of model variation is understandable owing to both the computational expense associated with running large suites of simulations, as well as the complexity associated with analyzing the results.
However, this same lack of model exploration makes it challenging to interpret observational results due to a lack of accurate uncertainties from assumptions in the models.

An additional challenge lies in generating large-enough statistical samples of high-resolution MW-mass galaxies to adequately discern how satellite properties depend on host    environmental differences or formation histories.
For example, a significant portion of the halo-to-halo variance in satellite statistics can be traced to differences in the formation histories of the host DM halos, such as their accretion timelines and merger events, which shape the resulting satellite distributions. 
Previous studies have established a negative correlation with the stellar mass of the central galaxy at fixed halo mass \citep{Anbajagane2020stellar,Farahi2020Aging}, which is due to the fact that more massive centrals reside in earlier-forming halos \citep{Zehavi2018Assembelybias,Lyu2023AssemblyBias}, where satellites experience stronger tidal stripping, dynamical friction, and longer exposure to environmental processing, ultimately lowering their surviving satellite counts.

The large dynamic range of low-mass satellites orbiting MW-mass hosts makes these systems computationally expensive to simulate, resulting in a small sample of MW-mass hosts with well-resolved satellite populations.  Suites of zoom-in simulations of MW-mass halos have accelerated such studies, but have been limited in the overall number of host galaxies that can be generated, usually on the order of $\mathcal{O}(10)$~\citep[e.g.,][]{Sawala_2016,Grand_2017,Wetzel_2023}.  These include N-body suites such as Symphony~\citep{symphony} and MW-est~\citep{2024Buch}.  The latter in particular has allowed for a deeper investigation into more rare formation histories, such as the accretion of satellites as massive as the Large Magellanic Cloud~(LMC), that provide a better sense of how the satellites around the MW may differ from other similar galaxies~\citep{2024Buch}. These N-body suites are complemented by hydrodynamic versions, including Numerical Investigation of a Hundred Astrophysical Objects~\citep[NIHAO;][]{Wang2015}, Feedback In Realistic Environments~\citep[FIRE;][]{FIRE1}, Auriga~\citep{Grand_2017}, and A Project Of Simulating the Local Environment~\citep[APOSTLE;][]{Sawala_2016}.

Uniform-box simulations can also play an important role, providing larger volumes with many MW-mass galaxies, albeit generated at lower resolutions than zoom ins.  One example is the TNG50 box~\citep{2019Nelson, 2019Pillepich} from the IllustrisTNG project~\citep{2018Pillepicha, 2017Weinberger, 2018Weinberger}, which contains 198 MW-mass galaxies and can resolve satellites down to $\mstar\gtrsim10^6~\mdot$~\citep{2024Pillepich}.
The spectrum of simulated systems include MW-mass galaxies that host diverse satellite populations that range from satellite-rich systems---like our own Galaxy---to nearly isolated ones~\citep{2024Pillepich}. 

Confronting these simulation challenges is critical as we move into an era of increasingly precise and extensive observational studies of nearby galaxy satellite populations. 
The most complete census of satellites is available for the MW and Andromeda~(M31) due to their proximity~\citep{2015Drlica-Wagner,2015Bechtol,2015Alam,2018Newton}.
These satellites provide a detailed picture of the formation of low-mass galaxies through their star formation histories~\citep[SFHs;][]{2014Skillman,2021Rusakov}, internal dynamics~\citep{2013Tollerud,2017Wheeler,2021Martinez-Garcia}, and chemical abundances~\citep{2006Munoz,2008Suda}.
In many cases, individual stars have been observed in these satellites~\citep{2015delPino,2017Karczmarek,2020McConnachie}, providing an unprecedented level of detail into the structure and history of these galaxies.
These observations offer a unique opportunity to compare with simulations and better understand the physics that governs galaxy formation and evolution at the smallest scales~\citep[for reviews, see][]{2022Sales,2025DolivaDolinsky}.

Beyond the MW and M31, multiple surveys have examined satellites around similar galaxies such as Exploration of Local VolumE Satellites~\citep[ELVES;][]{ELVES} and Satellites Around Galactic Analogs~\citep[SAGA;][]{SAGAI, SAGAIII}.
ELVES targeted 28 galaxies with host stellar masses between $10^{10}$ and $10^{11}~\mdot$ within 12 Mpc of the MW and observed satellites with $\mstar\gtrsim5\times10^5~\mdot$.
The SAGA Survey characterized the satellite populations around 101 MW-like galaxies, but down to a shallower satellite stellar mass of $\mstar>10^{7.5}~\mdot$ in its highest-confidence data. 
Taken together, data from surveys such as ELVES and SAGA  provide an unparalleled opportunity to understand the  distribution in satellite populations, as well as their internal properties.  The situation will only continue to improve with upcoming data from e.g., the Merian Survey~\citep{2025ApJ...993..110D}.  

The DaRk MattEr and Astrophysics with Machine learning and Simulations Project~\citep[DREAMS;][]{DREAMS}\footnote{\url{https://dreams-project.readthedocs.io}\\ \url{https://www.dreams-project.org}} provides thousands of cosmological hydrodynamic MW-mass zoom-in simulations with varied initial conditions and  galaxy-formation physics.  This approach is inspired by the Cosmology and Astrophysics with MachinE Learning Simulations~\citep[CAMELS;][]{CAMELS,CMD}, but is focused on small-scale structure observables.   
In this work, we focus specifically on the new DREAMS simulation suite for a Lambda Cold Dark Matter~($\Lambda$CDM) cosmology, introduced in Rose~et~al.~(submitted).  The suite includes 1,024 MW-mass halos, each run with distinct choices of astrophysical model parameters within the TNG galaxy-formation framework and cosmological variations.  These simulations provide an unprecedented opportunity to quantify uncertainties in the modeling of baryonic physics compared to intrinsic halo-to-halo variance in a range of satellite properties. We focus specifically on the  satellite mass function~(SMF), radial distribution, internal dark matter~(DM) density profile, and stellar half-light radius, comparing the results to observations from SAGA where available.

The structure of this paper is as follows. Section~\ref{sec:methods} briefly reviews the DREAMS simulations and describes the emulated datasets that are used in this analysis.  It also defines the four key satellite properties studied in this work. Sections~\ref{sec:population} and \ref{sec:individual} then present the results for the satellite population and internal properties, respectively. 
Section~\ref{sec:discussion} summarizes the results and discusses directions for future study. Appendix~\ref{app:nehod} describes the emulator procedure and accompanying validation tests.  Appendix~\ref{app:resolution} describes relevant resolution tests.

\section{Methods}
\label{sec:methods}

This paper utilizes the DREAMS CDM suite of MW-mass galaxies, which is summarized in Table~\ref{tab:DREAMS}.  This suite was first introduced in~Rose~et~al.~(submitted)\footnote{A preprint of this work was published on arXiv on Dec 2 2025 and is expected to appear as Rose~et~al.~(2025).} and we refer the interested reader to that work, as well as \citet{DREAMS}, for complete details on the simulation procedure and justification for the parameter variations.  Section~\ref{sec:sim_over} briefly highlights relevant aspects of the simulation setup for the analyses performed here.  Section~\ref{sec:weights} reviews the procedure by which certain regions of parameter space are down-weighted if they produce unrealistic galaxies.  Section~\ref{sec:calcs} describes how the satellite properties analyzed throughout this work are obtained from the simulations, while Section~\ref{sec:nehod} describes how the emulated samples used in the analyses are generated.

\begin{table*}
    \centering
    \renewcommand{\arraystretch}{1.1}
    \begin{tabular}{l|lccc|cccccc}
    \Xhline{3\arrayrulewidth} 
        DREAMS Suite    & Type   &   ICs   & $M_{\rm DM}/\mdot$   & $M_{\rm b}/\mdot$      & \multicolumn{2}{c}{Cosmology Parameters}   & \multicolumn{3}{c}{Astrophysics Parameters}\\
        &  &  &    $\times(\Omega_{\rm m} / 0.314)$    &   &   $\Omega_{\rm m}$ & $\sigma_8$           & $\widebar{e}_w$ & $\kappa_w$ & $\epsilon_{f, {\rm high}}$ \\
        \hline
        CDM MW    & N-body   & 1024 &  $2.0\times10^6$   & $0$  &  [0.274, 0.354]  & [0.780, 0.888] & --- & --- & --- \\
                  & TNG  & 1024       & $1.8\times10^6$    & $2.8 \times 10^5$ &  [0.274, 0.354]  & [0.780, 0.888] & [0.9, 14.4] & [3.7, 14.8] & [0.025, 0.4]\\
    \Xhline{3\arrayrulewidth}     
    \end{tabular}
    \caption{Summary of the DREAMS MW-mass simulation suites. The first is N-body, while the second is hydrodynamical, adopting the TNG galaxy-formation model.  Each suite contains 1,024 MW-mass galaxies, each with distinct initial conditions.  The table lists the simulation particle mass for DM~($M_{\rm DM}$) and baryons~($M_{\rm b}$), as well as the parameter ranges considered for five other parameters.
    The DM particle resolution depends on $\Omega_{\rm m}$ and is thus different in every simulation.
    The parameters include two cosmological parameters ($\Omega_{\rm m}$ and $\sigma_8$) and three astrophysical parameters that affect the SN wind energy~($\widebar{e}_w$) and speed~($\kappa_w$), as well as AGN energy~($\epsilon_{f, {\rm high}}$). See Section~\ref{sec:sim_over} for the definitions of these parameters.} 
    \label{tab:DREAMS}
\end{table*}
\subsection{Overview of Simulations}
\label{sec:sim_over}

The simulations are generated using the moving mesh code \textsc{arepo}~\citep{Springel2010, Springel2019, Weinberger2020} with the TNG galaxy-formation model~\citep{2018Pillepicha, 2018Weinberger}.
The initial conditions for the zoom-in simulations are created using \textsc{music}~\citep{2011Hahn}, following the procedure outlined in Appendix~A of~\cite{DREAMS} and with some updates presented in Section~2.2 of Rose~et~al.~(submitted). They uniformly cover a halo mass\footnote{The halo mass is defined as the sum of all matter within $R_{200}$, the radius at which the mean density of the halo is $200$ times the critical density of the universe.} range between $(5$--20$)\times10^{11}~\mdot$. To improve computational time, we impose an isolation criterion requiring that no other halo be more massive than $1.0\times10^{12}~\mdot$ within 1~Mpc of the target halo.
As a result, the galaxies in this suite are not true MW analogs, as they do not have a neighbor similar to M31.  

The DM particle mass resolution is $1.8\times(\Omega_{\rm m}/0.314)\times10^{6}~\mdot$, where \om\,is the total matter density, and the baryon particle mass resolution is $2.8\times10^5~\mdot$. The gravitational softening has a maximum physical length of $\epsilon_{\rm grav} = 0.441$~kpc. Galaxies are identified with the \textsc{subfind} algorithm~\citep{2001Springel} where gravitationally bound particles are assigned to 3D overdensities. We consider galaxies with stellar mass $\mstar>10^{7.5}~\mdot$ to be resolved in this analysis, but discuss where results may be affected by this choice throughout and in more detail in Appendix~\ref{app:resolution}.

A key advancement of the DREAMS suites is that they span uncertainties in both halo-to-halo variance and other model variations.  To this end, each simulation in the DREAMS CDM suite is generated from a unique initial condition, as well as a unique choice of five varied parameters.  Three of these parameters are astrophysical and are related to the feedback processes within the TNG model: two vary the energy and speed of the supernova~(SN) winds and one varies the energy output from Active Galactic Nuclei~(AGN) feedback. 

The SN parameters are related through the mass-loading factor, defined as the ratio of the mass outflow rate to the star formation rate~(SFR),
\begin{equation}
    \label{eq:mass_loading}
    \eta_w = \frac{2}{v_w^2} e_w \left(1 - \tau_w \right) \, ,
\end{equation}
where $v_w$ is the wind velocity, $e_w$ is the specific energy of the winds in units of $10^{51}~\text{erg}~\mdot^{-1}$, 
and $\tau_w$ is the fraction of energy released thermally.
The latter is fixed to the TNG fiducial value of $\tau_w = 0.1$.  We vary $\widebar{e}_w$, a unitless multiplicative factor that is directly proportional to $e_w$;  it is varied logarithmically about its fiducial TNG value of 3.6 within the range $\widebar{e}_w\in[0.9, 14.4]$.

The second astrophysical parameter varied in this suite, $\kappa_w$, is a unitless normalization factor that can impact the speed of SN winds, defined as
\begin{equation}
    v_w = \text{max}\left[ \kappa_w \sigma_{\scriptscriptstyle{\rm DM}} \left( \frac{H_0}{H(z)} \right)^{1/3}, v_{w, {\rm min}} \right] \, ,
    \label{eq:speed}
\end{equation}
where $\sigma_{\scriptscriptstyle{\rm DM}}$ is the local DM velocity dispersion, $H(z)$ is the Hubble constant (with $H_0$ the present-day value), and $v_{w, {\rm min}} = 350~\text{km/s}$ is the minimum wind speed. 
The local velocity dispersion is computed from the 64 nearest neighbors to the star particle producing the SN.
$\kappa_w$ is varied logarithmically between $\kappa_w \in [3.7, 14.8]$ with the TNG fiducial value being  7.4.  Note that $\kappa_w$ is less effective in low-density areas and at low redshift where the speed is simply set to $v_{w, {\rm min}}$.
 
The final feedback parameter considered, $\epsilon_{f,\mathrm{high}}$, sets the fraction of AGN energy that is transferred to the nearby gas.
The overall energy from AGN in the high-accretion state, which is most common to MW-mass galaxies, is given by 
\begin{equation}
    \dot{E}_{\text{AGN}}  = \epsilon_{f,\mathrm{high}} \,  \epsilon_r \, \dot{M}_{\scriptscriptstyle{\rm BH}} \, c^2 \, ,
\end{equation}
where $\epsilon_r$ is the radiative efficiency,  $\dot{M}_{\scriptscriptstyle{\rm BH}}$ is the black hole~(BH) accretion rate, and $c$ is the speed of light.
Halos are seeded with a BH of mass $M_\mathrm{BH}=1.2\times10^6~\mdot$ once their total mass exceeds $7.2\times10^{10}~\mdot$.
Therefore, most satellites that orbit MW-mass galaxies will never host a BH in the TNG model.
The fiducial value for $\epsilon_{f,\mathrm{high}}$  within TNG is 0.1 and we vary the parameter logarithmically such that $\epsilon_{f,\mathrm{high}} \in [0.025, 0.4]$.

In addition to the three astrophysical parameters described above, two cosmological parameters are uniformly varied: $\Omega_{\rm m}\in[0.274,0.354]$  and $\sigma_{\rm 8}\in[0.780, 0.888]$.
These ranges correspond to 2$\sigma$ uncertainties on the 2013 Planck temperature results~\citep{2014Planck}, and contain the Planck 2016 values with which the original TNG simulations were run \citep{2016Planck,2018Pillepicha}.
The ranges of cosmological parameters are intentionally kept larger than the current Planck constraints to minimize prior effects on our results.
All other cosmological parameters, $h = 0.691$ and $\Omega_{\rm b} = 0.046$, are fixed for each simulation and are consistent with Planck 2016 values~\citep{2016Planck}.

In total, the hydrodynamic suite comprises 1,024 simulations, each with a unique initial condition and choice of parameters \citep[selected following a Sobol sequence; ][]{sobol}.  For each hydrodynamical simulation, there is a corresponding N-body simulation of the same galaxy at the same resolution\footnote{The DM particle masses in Table~\ref{tab:DREAMS} differ because the N-body simulations treat all matter as collisionless, subsuming the baryonic mass into the DM particle.}.  The N-body suite has the same cosmological parameter variations as its sister suite.  

\begin{figure*}
    \centering
    \includegraphics[width=\textwidth]{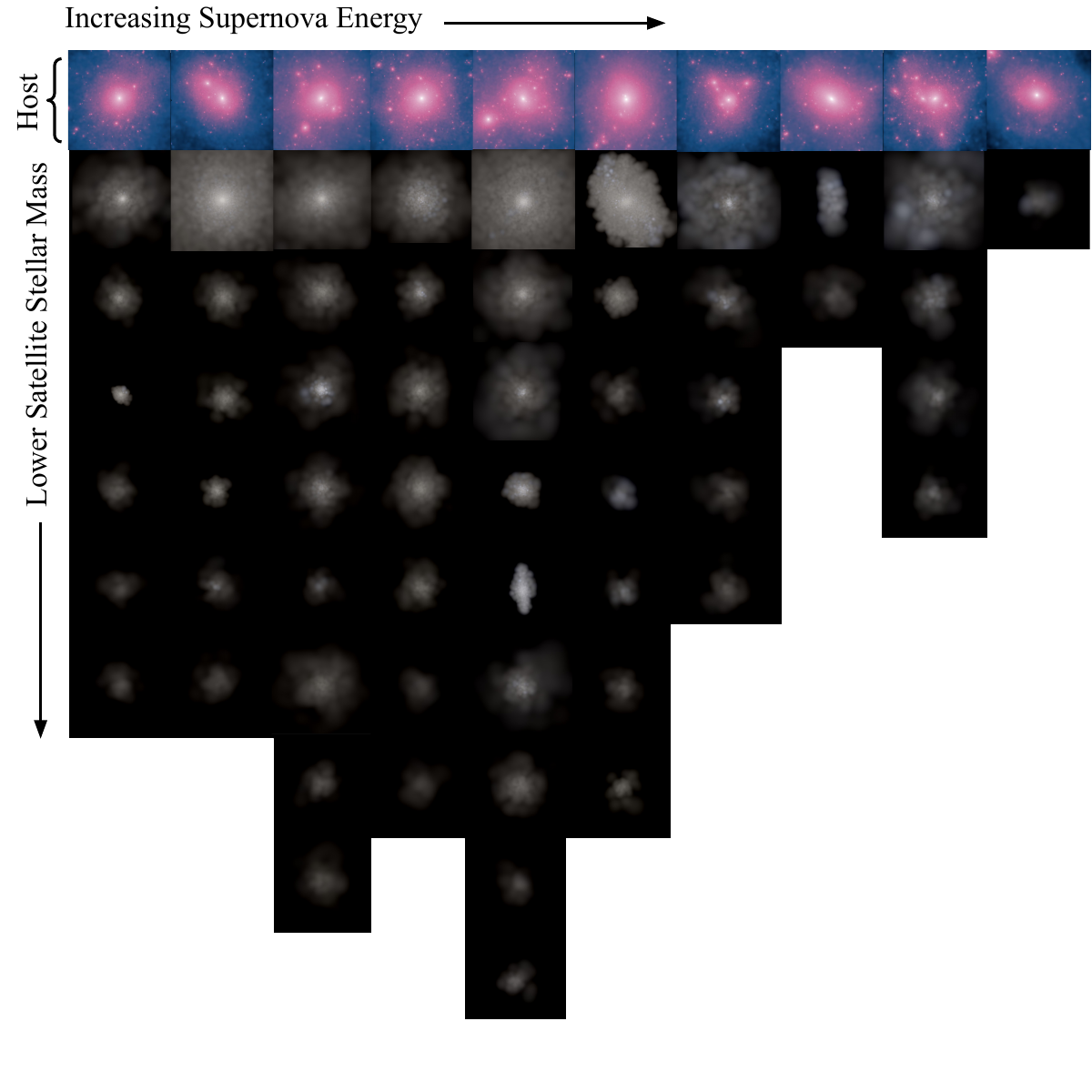}
    \caption{A gallery of the first 10 (out of 1,024) simulated MW-mass systems from the DREAMS CDM suite.
    The top row displays DM density projections for the host and its surrounding satellites out to 300~kpc, ordered from left to right by increasing SN wind energy, \ew.
    Each column below the top row shows the corresponding population of resolved satellite galaxies ($\mstar>10^7~\mdot$) for that host.
    The satellite images are shown in order from most to least massive and have a $20\times20$~kpc field of view.
    The red, green, and blue color channels for the satellite images are created from the $i$, $r$, and $g$ Sloan filter luminosities.
    The figure highlights the significant variation in satellite numbers, sizes, and overall appearance.
    }
    \label{fig:sat_images}
\end{figure*}

Figure~\ref{fig:sat_images} displays images of the first 10 boxes from the hydrodynamic DREAMS suite with the DM density for each host shown in the top row.  Below each host, we show the stellar light projections for its resolved satellites.
The columns are ordered by increasing \ew\,to highlight the satellite population's dependence on SN feedback, mainly seen through the decreasing number of satellites with increasing SN feedback, as well as the significant differences from halo-to-halo variance.

\subsection{Weighting Procedure}
\label{sec:weights}

A key feature of the DREAMS suites is that they include MW-mass galaxies that are simulated under different combinations of cosmological and astrophysical parameters.  This provides a unique opportunity for exploring how parameter variations impact the properties of the MW host as well as its satellites.  However, it also raises the question of how to present a result---such as the predicted distribution for some observable of interest---when some of the generated hosts may be inconsistent with observed galaxies. Although the range for each parameter described in Section~\ref{sec:sim_over} is chosen to align with current uncertainties, when all five parameters are varied simultaneously, it is possible to produce unrealistic galaxy populations. 
To this end, we assign weights throughout the 5D parameter space based on how well MW-mass galaxies simulated with those parameters reproduce observed scaling relations.
This procedure allows one to down-weight the parts of parameter space that are unphysical. 
We summarize the procedure here for reference---see Rose~et~al.~(submitted) for further details.

The weights are calculated with a `sudo-posterior' method based on how well the central MW-mass galaxy, simulated with a particular set of parameters $\boldsymbol{\theta}$, reproduces the `UM-SAGA' stellar mass--halo mass relation from \cite{SAGAV}. For this procedure, we fit the relation with a piecewise linear function $g(X)$, where the slope/intercept can depend on $X$, the halo mass. 

We consider a grid of unique parameter combinations of \kw, \ew, and \agn.\footnote{Cosmological parameters are not included in the weighting scheme to reduce computation time, as they have minimal effect on the central galaxy properties.}  For bin $j$ in the grid, we emulate $N_j = 10^3$ galaxies, predicting their stellar and halo mass.  These outputs are respectively indicated as $X_{j,n}^{\mathrm{em}}$ and $Y_{j,n}^{\mathrm{em}}$ for the $n^{\rm th}$ galaxy in the bin.
We then compare the entire batch of $N_j$ samples with the observed scaling relation by calculating the residual 
\begin{equation}
    R_{\theta_j} = \frac{1}{N_j} \sum_{n=1}^{N_j} \left(Y_{j,n}^{\rm em} - g\left(X_{j,n}^{\rm em}\right) \right) \, .
\end{equation}
A zero residual corresponds to a parameter combination, $\theta_j$, that produces galaxies near the mean of the observed scaling relation.

The residual is then converted into a weight, $\widetilde{w}_j$, using a Gaussian kernel with a free temperature parameter, $\tau$,
\begin{equation}
   \widetilde{w}_j = \mathrm{exp} \left( - \frac{R^2_{\theta_j}}{2 \tau ^2} \right).
\end{equation}
The normalized weight is defined as $w_j = \widetilde{w}_j/\sum_j \widetilde{w}_j$. 
The free parameter, $\tau$, is a measure of uncertainty on the mean observed relation.
Based on the uncertainties discussed in~\citet{2019Behroozi}, $\tau$ is set to 0.2~dex to account for variation between other predicted SMHM relations (see their Figure~34) and potential systematic uncertainties.

The weight for a specific parameter combination quantifies how well a galaxy-formation prescription matches observations. 
The weights are applied for all results presented in this work where all three astrophysical parameters are varied.
The main conclusions of the paper are not sensitive to the weighting scheme, although the parameter means and standard deviations may shift slightly.  In general, repeating the analysis with no weights does not change the primary results.  

Currently, the parameter weights only account for one observed scaling relation, the present-day SMHM relation.  
Their constraining power can be updated in the future by including other scaling relations such as the mass-metallicity~\citep{2020Curti} or the BH mass to velocity dispersion~\citep{2020Greene} relation, as well as redshift-dependent scalings.

\subsection{Satellite Property Calculations}
\label{sec:calcs}

A satellite about a simulated MW-mass host is selected if its 3D radial distance from the minimum of the host potential is less than 300~kpc, its stellar mass is greater than $10^{7.5}~\mdot$, its total mass is greater than $10^8~\mdot$, and the satellite is a member of the host Friends-of-Friends~(FoF) group\footnote{This is in contrast to the SAGA selection criteria which uses a satellite velocity selection, relative to the host, of $v_{\rm sat}<275$~km/s~\citep{SAGAIII}.} unless otherwise stated. In general, the stellar mass for a satellite is the summed mass of its bound star particles, as determined by \textsc{subfind}. For the satellite populations across all DREAMS hosts, we study four key properties: the SMF, the 2D radial distribution~(2D $R_\mathrm{sat}$), the DM density slope~($\alphaslope$), and the 2D half-light radius~(2D $R_\mathrm{half}$).  This subsection defines each of these properties in turn.

The SMF quantifies the average number of satellites per host galaxy, $N_{\rm sat}$, as a function of the satellite's stellar mass, and is computed as a differential quantity, $dN_\mathrm{sat}/d\mathrm{log}M_*$. In practice, we define a series of $k$ equal-spaced bins of width $\mathrm{log} M_*$ in stellar mass, spanning the range log($M_*/\mdot)=[7.5,10]$.
Then, for the $i^{\rm th}$ host, we count the number of satellites, $N_{i,k}$, whose stellar mass falls within bin $k$. The differential SMF for an individual host, $\Phi_{i, k}$, is defined as 
\begin{equation}
    \Phi_{i,k} = \frac{N_{i,k}}{\Delta \mathrm{log} M_*} \, .
\end{equation}
This provides a distinct SMF for each host system in the simulation suite.

To determine the mean SMF for a collection of hosts, we calculate the weighted average of all individual host SMFs.  Then, the mean SMF for the $k^{\rm th}$ stellar-mass bin, $\langle \Phi_k \rangle$, is given by
\begin{equation}
    \langle \Phi_k \rangle = \frac{\sum^{N_{\rm host}}_{i=1} w_i \Phi_{i,k}}{\sum^{N_{\rm host}}_{i=1} w_i} \, ,
\end{equation}
where $w_i$ is the weight for the $i^{\rm th}$ host and $N_{\rm host}$ is the total number of hosts considered.\footnote{Because each host is generated for a distinct combination of astrophysical parameters, it is equivalent to label it with the index $j$ associated with the parameter combination $\theta_j$ or index $i$ associated with the host number.} The 1$\sigma$ scatter is computed as the weighted standard deviation of the individual SMFs around this mean, where the standard deviation is normalized with an unbiased estimator of the covariance matrix defined in \cite{2022Farahi},
\begin{equation}
    \sigma_k = \sqrt{ A \sum^{N_{\rm host}}_{i=1} w_i \left(\Phi_{i,k} - \langle \Phi_k \rangle\right)^2} \,,
\end{equation}
with 
\begin{equation}
    A = \frac{\sum_{i=1}^{N_{\rm host}} w_i}{\left(\sum_{i=1}^{N_{\rm host}} w_i\right)^2 - \sum_{i=1}^{N_{\rm host}} w_i^2 } \,. 
\end{equation}

The second property, 2D $R_\mathrm{sat}$, describes the concentration of satellites as a function of projected distance from the center of the host.
For this case specifically, the satellites are not restricted to a physical 3D radius of 300~kpc, but have a looser constraint of 400~kpc.
For each host $i$, we obtain the 2D projected positions $(x,y)$ of its satellites relative to the minimum of the host's potential.
There is no rotation applied to the host, so the orientation of the satellites is random.
The projected radial distance for each satellite is calculated as $R_\mathrm{sat}=\sqrt{x^2+y^2}$.

We consider all satellites with a projected radius between 10-150~kpc to reduce the contamination from interlopers in the observational data~\citep[for a discussion, see][]{SAGAIII}.  The total number of satellites for host $i$ within this projected radius is $N_{\mathrm{sat},i}$ and they are then sorted by $R_\mathrm{sat}$.
The cumulative fraction for this host is 
\begin{equation}
    f_i(R_{\rm sat}) = \frac{N_{\mathrm{sat},i} (< R_{\rm sat})}{N_{\mathrm{sat},i}} \, .
\end{equation}

To enable averaging across all hosts, we define a common set of $\ell$ radial bins, $R_{\rm sat, \ell}$, evenly spaced from 0 to 150~kpc.
We interpolate each host's cumulative fraction, $f_i(R_{\rm sat})$, onto these bins, $R_{\rm sat, \ell}$, resulting in a set of values $f_{i,\ell}$. The mean normalized radial distribution, $\langle f_\ell \rangle$, and its 1$\sigma$ scatter, $\sigma_\ell$, are then computed using the same scheme as described for the SMF,
\begin{align}
    \langle f_\ell \rangle &= \frac{\sum^{N_{\rm host}}_{i=1} w_i f_{i,\ell}}{\sum^{N_{\rm host}}_{i=1} w_i} \\
    \sigma_\ell &= \sqrt{ A \sum^{N_{\rm host}}_{i=1} w_i \left(f_{i,\ell} - \langle f_\ell \rangle \right)^2 } \,.
\end{align}

The third property, $\alphaslope$, describes the slope of the spherically averaged DM density profile at (physical radius) 1.5~kpc.
The density is computed using a method analogous to standard smoothed-particle hydrodynamics~(SPH).
The DM particles that are bound to a given satellite are selected by \textsc{subfind}.
To measure the density at a specific radius $r$, we distribute 100 sample points evenly on the surface of a Fibonacci sphere~\citep{fibonacci} of that radius.
For each of these sample points, we find the 32 nearest neighboring DM particles using a k-d tree.
The smoothing length, $h_{\rm sph}$, is defined as the distance to the $32^{\rm nd}$ neighbor, and the enclosed mass, $M_\mathrm{enc}$, is the sum of the masses of those 32 particles.
The density at that sample point is $\rho=M_\mathrm{enc} / \left(\frac{4}{3} \pi h_\mathrm{\rm sph}^3 \right)$.
The final density at radius $r$, $\rho(r)$, is the average density of all 100 sample points.

To calculate the slope, $\alphaslope$, we compute this average density at two radii: $r_{\rm in}=1.3$~kpc and $r_{\rm out}=1.7$~kpc.
The slope is then determined using a finite difference in log space,
\begin{equation}
    \alphaslope = \frac{\mathrm{log} \, \rho(r_{\rm in}) - \mathrm{log} \, \rho(r_{\rm out}) }{\mathrm{log} \, r_{\rm in} - \mathrm{log}\, r_{\rm out}} \,.
\end{equation}
Section~\ref{sec:dm} analyzes the relationship between the inner slope and the total satellite mass, $M_\mathrm{TOT}$, for galaxies in the DREAMS N-body suite. 
For a bin in log~$M_\mathrm{TOT}$, we calculate the mean value of the inner slope, $\langle \alphaslope \rangle$, and the standard deviation.  Galaxies taken from the N-body suite do not have weights because they have no SN or AGN feedback.  These statistics represent the average slope-mass relation and intrinsic 1$\sigma$ scatter of the satellite population around that mean relation.

Finally, we analyze the satellite size-mass relation by measuring the 2D stellar half-light radius, 2D $R_\mathrm{half}$, as a function of $M_*$.
The $R_\mathrm{half}$ value for each satellite is determined from a 2D luminosity map, which is generated using an SPH-like smoothing technique. Specifically, we select all star particles within a field of view set to 3$\times$ the satellite's 3D stellar half-mass radius as defined by \textsc{subfind}.
We then take the r-band luminosity~\citep{2010Doi} for each particle from the \textsc{subfind} catalog and smooth it using a kernel defined by its 32 nearest neighbors~\citep{2015Torrey, 2018Nelson}. To obtain a 2D luminosity density image, we 
 project the smoothed luminosity onto a 512$\times$512 pixel grid. This image is used to measure the half-light radius.
We use \textsc{photutils}~\citep{photutils,astropy:2013,astropy:2018,astropy:2022} to measure the properties of the satellite, including its centroid, position angle, and ellipticity $\epsilon$. The axial ratio is defined as $q=1-\epsilon$.

The satellite's total luminosity, $L_\mathrm{TOT}$, is measured within a large elliptical aperture set to 5$\times$ the satellite's semi-major axis. We then iteratively determine the effective semi-major axis, $a_{\rm eff}$, of an elliptical aperture that shares the same centroid, position angle, and axial ratio, enclosing exactly half of the total luminosity.
This semi-major axis is `circularized' to find the half-light radius,
\begin{equation}
    R_\mathrm{half} = a_\mathrm{eff} \sqrt{q} \,.
\end{equation}
This circularized half-light radius is converted from pixels to physical kpc using the pixel scale, defined as the field of view in kpc divided by the number of pixels (512).

With the $R_\mathrm{half}$ and $M_*$ values computed for all satellites, we bin the population by the logarithm of $\mstar$.
To ensure consistency with the SMF and radial distribution statistics, we first calculate the mean half-light radius, $\widebar{R}_{i,k}$, for the satellites belonging to host $i$ within mass bin $k$.
The weighted mean half-light radius for the $k^{\rm th}$ bin is then calculated as the average of these host means, considering only hosts that contain satellites in that bin,
\begin{equation}
    \langle R_{\mathrm{half},k} \rangle = \frac{\sum_{i \in S_k} w_i \widebar{R}_{i,k}}{\sum_{i \in S_k} w_i} ,
\end{equation}
where $S_k$ represents the set of hosts that possess at least one satellite in mass bin $k$.
The 1$\sigma$ intrinsic scatter is the weighted standard deviation within that bin, given by
\begin{equation}
    \sigma_{k} = \sqrt{ A_k \sum_{i \in S_k} w_i \left(\widebar{R}_{i,k} - \langle R_{\mathrm{half},k} \rangle \right)^2 } .
\end{equation}

For plotting on a logarithmic scale, the 1$\sigma$ scatter is represented by asymmetric error bars, where the relative magnitude of the intrinsic scatter is kept constant.
Here, the upper bound is set to $\langle R_\mathrm{half} \rangle + \sigma$ and the lower bound is set to $\langle R_\mathrm{half} \rangle^2 / (\langle R_\mathrm{half} \rangle + \sigma)$.

\subsection{Emulation of Satellite Galaxies}
\label{sec:nehod}

\begin{table*}
    \centering
    \renewcommand{\arraystretch}{1.1}
    \hspace{1cm}
    \begin{tabular}{lccccccc}
    \Xhline{3\arrayrulewidth} 
        Emulated Dataset   &  \# of MWs      & \multicolumn{2}{c}{Cosmology Parameters}   & \multicolumn{3}{c}{Astrophysics Parameters}       \\
                        &                 & $\Omega_{\rm m}$ & $\sigma_8$              & $\widebar{e}_w$ & $\kappa_w$  & $\epsilon_{f, {\rm high}}$ \\
        \hline
        DREAMS Fiducial  & $10^5$          &  0.314            & 0.834                   & 3.6             & 7.4         & 0.1 \\
        DREAMS Varied      & $10^5$          &  [0.274, 0.354]  & [0.780, 0.888]          & [0.9, 14.4]     & [3.7, 14.8] & [0.025, 0.4]\\
        $\widebar{e}_w$ Varied & $5\times10^5$ &  0.314     & 0.834            & \{0.9, 1.8, 3.6, 7.2, 14.4\} & 7.4         & 0.1 \\
        DREAMS N-body          & $10^5$          &  0.314            & 0.834                   & N/A             & N/A         & N/A \\
         \Xhline{3\arrayrulewidth}    
    \end{tabular}
    \caption{Summary of the emulated datasets that are used in this work.  The first three datasets are generated by training the NeHOD emulator~\citep{NeHOD} on the hydrodynamical CDM DREAMS suite, while the last is obtained by training on the N-body suite.  Depending on the dataset, some cosmological and astrophysical parameters are fixed to their fiducial values from TNG, while others are varied. Each dataset consists of $10^5$ samples; note that `\ew Varied' is the combination of five datasets, each with a different value of \ew.  }
    \label{tab:NeHOD}
\end{table*}

While the DREAMS CDM suite contains an unprecedented number of cosmological MW zoom-in simulations, it still covers only a subset of the possible combinations of the five free parameters considered.  Ideally, one would interpolate over this discrete sampling to better understand how the properties of the simulated galaxies depend on the parameter variations.  Toward this end, we use the DREAMS suite to train an emulator that produces a population of satellite galaxies for a MW-mass host for any specified choice of $\Omega_{\rm m},~\sigma_8,~\widebar{e}_w,~\kappa_w,~\text{and}~\epsilon_{f, {\rm high}}$.  The efficiency of this approach allows one to populate satellites in orders-of-magnitude more host galaxies.

In particular, we utilize the NeHOD emulator~\citep{NeHOD}, which is a generative model that combines a normalizing flow~\citep{2016Rezende,2017Papamakarios,2019Papamakarios} and a Variational Diffusion Model (VDM)~\citep{2021Vahdat,2021Song,2023Kingma} with a Transformer noise prediction model~\citep{2023Vaswani} to create new realizations of galaxies and their satellites from a suite of simulations. 
Training on the DREAMS Warm Dark Matter MW-mass suite from~\cite{DREAMS}, \cite{NeHOD} showed that NeHOD reproduces complex properties of satellites around MW-mass hosts with varied input simulation parameters.  The results presented in this work utilize the NeHOD emulator by retraining it on the new CDM suites presented here.  Details of the NeHOD emulator, as well as validation tests, are provided in Appendix~\ref{app:nehod}, and the results are summarized here.

Training NeHOD is done in two steps.
The first step is to train the normalizing flow to reproduce properties of the central MW-mass galaxy given the simulation parameters used to create it.
For this analysis, we train the normalizing flow on the five simulation parameters (\om, \s8, \ew, \kw, and \agn) to predict three properties of the central halo:~(i)~the number of satellites with a stellar mass above $10^6~\mdot$ and a total mass above $10^7~\mdot$,
(ii)~the stellar mass of the central galaxy defined as all stellar matter within a sphere of radius equal to $2\times$ the 3D stellar half-mass radius,
and (iii)~the total mass of the system within $R_{200}$.\footnote{The cuts that are used to select satellites for NeHOD training are less restrictive than those used in other parts of the analysis. The reason for this is to avoid data near the edge of NeHOD's training dataset, which may be affected by the sharp cut imposed on the satellite mass function.}  
Note that satellites that have been severely stripped of DM such that $\mstar>10^6~\mdot$ and $M_\mathrm{TOT} < 10^7~\mdot$ are removed from the sample; these severely stripped galaxies will likely be missed by \textsc{subfind}, which requires at least 20 DM particles in a subhalo. 

The second NeHOD step is to train a VDM that is used to produce the satellite properties. 
The VDM takes as input the same properties of the central galaxy produced as target parameters by the normalizing flow.
The satellites are then treated as a point cloud, where the properties of each are attributes of the points in the cloud. 
For this analysis, we focus on seven satellite properties for the NeHOD training.
The first three are the galactocentric positions of the satellite in the $x$, $y$, and $z$ directions.
The others are four satellite properties discussed in Section~\ref{sec:calcs}: the SMF, the 2D radial distribution, the DM density slope, and the 2D half-light radius.

Table~\ref{tab:NeHOD} summarizes the emulated datasets created in this work.  Each dataset consists of at least $10^5$ galaxies to densely sample the full 5D parameter space and produce smooth distributions of binned galactic properties.  However, the results presented here are not sensitive to this choice as long as the total dataset contains at least $10^4$ samples (see Appendix~\ref{app:nehod}).
The first dataset, `DREAMS Varied', varies the five simulation parameters uniformly across the span of each parameter in the DREAMS suite.
The second dataset,  `DREAMS Fiducial', has the three free astrophysical parameters fixed to the fiducial TNG values \citep[$\widebar{e}_w=3.6$; $\kappa_w=7.4$; $\epsilon_{f,\mathrm{high}}=0.1$;][]{2018Pillepicha} and the two cosmological parameters fixed to the fiducial Planck values~\citep[$\Omega_{\rm m}=0.314$; $\sigma_8=0.834$;][]{2014Planck}. 
The third dataset keeps all parameters except \ew~fixed at their fiducial TNG values and contains five subsets of data, each with $10^5$ galaxies, where each subset has a different value of \ew~that spans the range of its parameter space.
While we explore the variations in the other four parameters and discuss them throughout the text, we highlight variations to \ew~because they have the largest effect on satellite properties.
The fourth and final dataset, `DREAMS N-body', is produced from the N-body suite of simulations following the same procedure, varying only the two cosmological parameters.

Appendix~\ref{app:nehod} details various training validation tests and sample convergence tests.
In summary, hyperparameter tuning is done with \textsc{optuna}~\citep{optuna} and epistemic uncertainties are included in each dataset through emulator ensembling.
Through training on different numbers of simulations, we find minimal deviations if the training dataset is reduced to 640 simulations, but slight differences suggest that results could be improved by increasing the sample size of DREAMS galaxies beyond 1,024 simulations.
Additionally, the results converge once the emulated dataset consists of at least $\sim$$2\times10^3$ samples.

\section{Satellite Population Statistics}
\label{sec:population}

This section explores the properties of the satellite populations surrounding MW-mass galaxies, including their overall number and spatial distribution. 
Section~\ref{sec:massposition} presents the mass and spatial distribution for these satellites, and the results are then compared to observations in Section~\ref{sec:saga}.

\begin{figure*}
    \centering
    \includegraphics[width=\columnwidth]{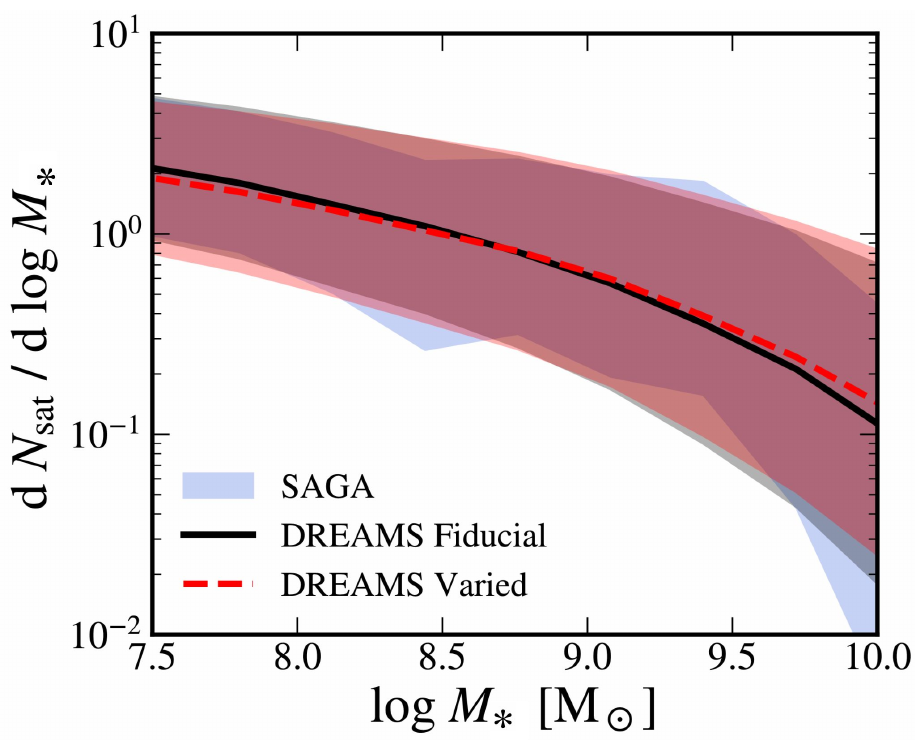}
    \includegraphics[width=0.98\columnwidth]{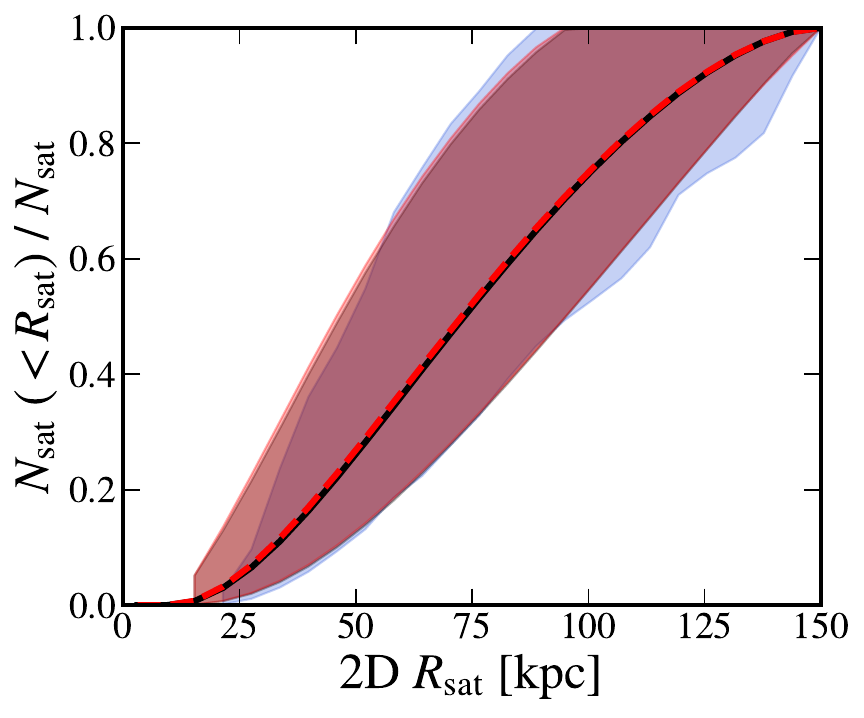}
    \caption{\textit{Left:} SMF for MW-mass galaxies in the emulated DREAMS Fiducial~(black) and Varied~(red) datasets, compared to the observations from SAGA~\citep{SAGAIII}.
    SAGA hosts are down-selected to fall within the mass range of the DREAMS host galaxies, reducing the observational sample from 101 to 80 MW-mass systems.  The satellites are taken from their Gold sample, which has $M_* \geq 10^{7.5}~\mdot$ and is 94\% complete. 
    \textit{Right:} Normalized 2D projected radial distribution, $R_{\rm sat}$, for the same satellites shown in the left panel. For both panels, lines indicate the average value, while the shaded bands show the 1$\sigma$ spread. 
    The emulated DREAMS data agrees well with the SAGA results for both distributions.  The primary exception is the radial distribution below 2D $R_{\rm sat} \lesssim 25$~kpc.  Notably, the DREAMS Fiducial and Varied results are comparable to each other, suggesting that uncertainties in the cosmological and astrophysical parameter variations are subdominant to intrinsic halo-to-halo variance and uncertainties in the host halo mass.
    }
    \label{fig:mass_func}
\end{figure*}

\vspace{1cm}

\subsection{Satellite Masses \& Positions}
\label{sec:massposition}

Figure~\ref{fig:mass_func} shows the satellite SMF~(left panel) and projected radial distribution~(right panel) for the MW-mass galaxies in the emulated DREAMS Fiducial~(black) and Varied~(red) datasets.  The solid line shows the mean value, while the band shows the $1\sigma$ spread.  Comparing these two datasets indicates the degree to which variations in the astrophysical and cosmological parameters affect the results.  The fact that the black and red lines/bands are consistent with each other for both the SMF and radial distribution suggests that the uncertainties are not dominated by variations to the cosmological and astrophysical parameters.  

To robustly contextualize these results, we note the three contributing sources to the total scatter of DREAMS Varied.
First, the emulated dataset spans a virial mass range of $(0.5\text{--}2.0)\times10^{12}~\mdot$ to account for the uncertainty in the MW's total mass and ensure that the host sample closely matches the mass distribution of the SAGA survey (see Section~\ref{sec:saga}).
Second, we incorporate theoretical uncertainties by varying five simulation parameters (\om, \s8, \ew, \kw, \agn), with the parameter-space constraints from Rose~et~al.~(submitted).
Finally, intrinsic halo-to-halo variance contributes from each of the 1,024 simulations being generated from a unique initial density field, leading to a unique formation history and environment.
We focus this analysis on the relative contribution of the five simulation parameters, as DREAMS is uniquely suited to probe these parameter variations, but also discuss the relative impact of halo-mass uncertainty.

The most massive satellites present in the dataset, $M_*>10^9~\mdot$, are not ubiquitous; the average number of these satellites is $<1$ around any MW host.
The existence of the LMC, which has a stellar mass of $\sim$$2\times10^9~\mdot$~\citep{2012Eskew}, around our own Galaxy highlights that the MW is a $\sim$$1\sigma$ deviation from average in this respect. 
The number of satellites with stellar masses above $10^9~\mdot$ does not change significantly from the DREAMS Fiducial to Varied datasets, increasing from $0.33^{+2.49}_{-0.29}$ to $0.36^{+2.64}_{-0.32}$.
For lower-mass satellites, $10^{7.5}<M_*<10^{9}~\mdot$, there is also little difference in the mean and spread of the number of satellites once parameter variations are introduced, with counts decreasing from $1.98^{+6.26}_{-1.50}$ to $1.86^{+6.41}_{-1.42}$.

We next examine how the SMF changes when each of the cosmological and astrophysical parameters is varied independently.
The largest change comes from variations in the SN wind energy (\ew), with a subdominant effect from the SN wind speed (\kw). There is almost no effect from variations to \om, $\sigma_8$, or the fraction of AGN energy transferred to nearby gas (\agn). 
The effect of each parameter is measured from a unique emulated dataset where that parameter is varied independently with the other parameters fixed to their fiducial values, similar to `\ew Varied' described in Table~\ref{tab:NeHOD}.
The lowest (highest) value of \ew~that we model produces $0.93^{+5.16}_{-0.79}$~($0.07^{+1.37}_{-0.00}$) satellites with $\mstar>10^9~\mdot$, and $3.44^{+10.05}_{-2.56}$~($1.00^{+5.16}_{-0.83}$) satellites with $10^{7.5}<\mstar<10^9~\mdot$. 
As also seen from the left panel of Figure \ref{fig:mass_func}, the variations in the SMF due to these physics variations are less than the variations from halo-to-halo variance and the uncertainty on the host halo mass.
The other SN parameter, \kw, produces a subdominant effect where the largest variations in parameter values result in $\sim$$15\%$ variations at the high-mass end and $\sim$$50\%$ variations at the low-mass end.
The other three parameters produce a change of $<20\%$.

Finally, we investigate the sensitivity of the satellite abundance to the total mass of the host halo.
We measure this by calculating the mean SMF and 1$\sigma$ spread for 3 groups of $5\times10^3$ host galaxies from the `DREAMS Fiducial' dataset whose host masses are closest to $5\times10^{11}~\mathrm{M_\odot}$, $1\times10^{12}~\mathrm{M_\odot}$, or $2\times10^{12}~\mathrm{M_\odot}$.
These datasets result in a spread of $\sim0.1$~dex about the target halo mass.
Since the simulation parameters are fixed in the `DREAMS Fiducial' dataset, this comparison targets the relative impact of the host mass and the intrinsic halo-to-halo variance.
We find that the normalization of the SMF is strongly dependent on the host mass, with the number of satellites scaling positively with the total mass of the system.
The smallest halo masses, $\mtot\sim5\times10^{11}~\mathrm{M_\odot}$, have $3\times$ fewer $\mstar=10^{7.5}~\mdot$ satellites and $2\times$ fewer $\mstar=10^{10}~\mdot$ satellites than the most massive hosts, $\mtot\sim2~\times~10^{12}~\mathrm{M_\odot}$.
However, the 1$\sigma$ spread about the mean SFM when the mass is fixed to $1\times10^{12}~\mathrm{M_\odot}$ remains the same as the full `DREAMS Fiducial' result.
Therefore, we find that while the SMF depends on both the SN wind energy and the host halo mass, the uncertainty on the prediction is dominated by the intrinsic halo-to-halo scatter.

The right panel of Figure~\ref{fig:mass_func} presents the 2D radial distribution of satellites around the central galaxy.
Similar to the SMF, the DREAMS Fiducial and Varied datasets predict nearly identical results, suggesting that halo-to-halo variance and host halo mass uncertainty dominate over uncertainties in the parameter modeling.
Across all radii, including parameter variations, produces $<1$\% deviations from the average radial profile with fiducial physics and cosmology.
For the DREAMS Fiducial~(Varied) dataset, there are $27.4^{+30.6}_{-14.5}$\%~($28.6^{+31.0}_{-14.9}$\%) of satellites within 50~kpc and $78.9^{+21.1}_{-19.8}$\%~($79.4^{+20.5}_{-19.5}$\%) of satellites within 100~kpc.
Neither the cosmological nor astrophysical parameters that are varied have a substantial effect on the satellite radial profile. 
The largest difference is between the extremes of the SN wind energy, \ew, where the average percentage of satellites within 100~kpc decreases from $79.7^{+20.3}_{-18.8}$\% to $77.4^{+22.6}_{-21.6}$\% going from lowest to highest $e_w$.
While the change is minimal, the difference likely comes from the lower average stellar mass of satellites when SN wind energy is high.

Unlike the SMF, the 2D radial distribution of satellites shows no significant dependence on the host halo mass.
For this distribution, the mean and 1$\sigma$ remain constant across the range of $(0.5-2.0)\times10^{12}~\mathrm{M_\odot}$, indicating that while a more massive host produces more satellites, their relative distribution within the halo remains unchanged.

It is worth highlighting that the distributions presented in Figure~\ref{fig:mass_func} can suffer numerical effects.  For example, the satellite stellar mass can increase with resolution (see Appendix~\ref{app:resolution}), which can change which satellites satisfy the $M_* > 10^{7.5}~\mdot$ selection.  This can shift the SMF or alter the radial distribution. Additionally, we expect some level of artificial disruption to affect satellites in the inner regions of the host~\citep{2018MNRAS.475.4066V}, which may worsen the discrepancy with SAGA data in this region.  Addressing both of these challenges requires a detailed study of how these distributions depend on resolution, which we save for future work.

\subsection{Comparisons to SAGA}
\label{sec:saga}

To contextualize the results from DREAMS, we compare them with the observed relations from the SAGA Survey, taken from the online database~\citep{SAGAIII}.  For a fair comparison to the DREAMS data, we remove any SAGA host galaxies with a total mass above $2.0 \times 10^{12}~\mdot$ and below $5 \times 10^{11}~\mdot$, resulting in a subsample of 80 galaxies. We also limit the analysis to SAGA's `Gold sample,' which corresponds to stellar masses $M_* \geq 10^{7.5}~\mdot$ and is 94\% complete. 
The SAGA Survey suffers from some contamination, `interlopers', which are field galaxies that are further than 300~kpc from their host in real space, but less than 300~kpc once projected in 2D.
These interlopers are expected to constitute up to $30\%$ of the SAGA satellites, dropping to $15\%$ within a projected distance of 150~kpc~\citep{SAGAII}.  As a result, we limit the comparison of the radial distribution profile to satellites falling within 2D~$R_{\rm sat} < 150$~kpc.
For the SMF comparison, we leave the radial selection as 3D~$R_\mathrm{sat}<300$~kpc, but find that the results are not sensitive to a 2D or 3D radial selection.

\begin{figure*}
    \centering
    \includegraphics[width=.96\columnwidth]{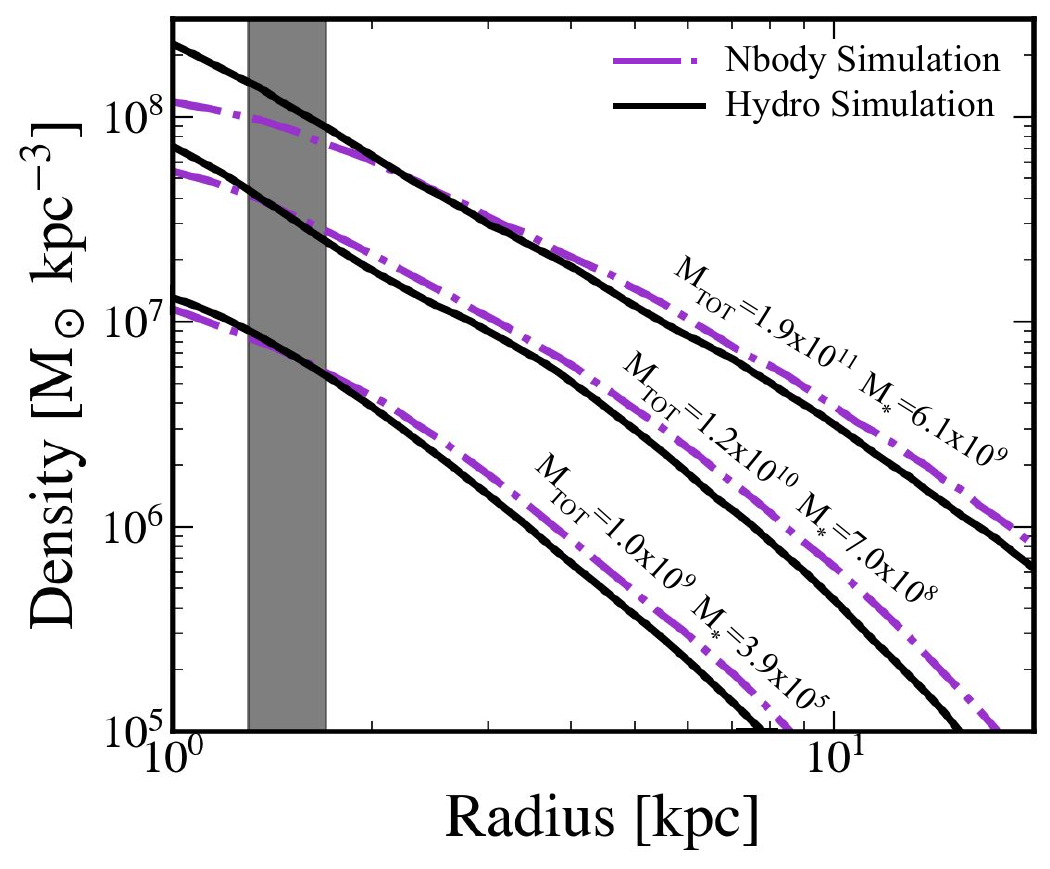}
    \includegraphics[width=\columnwidth]{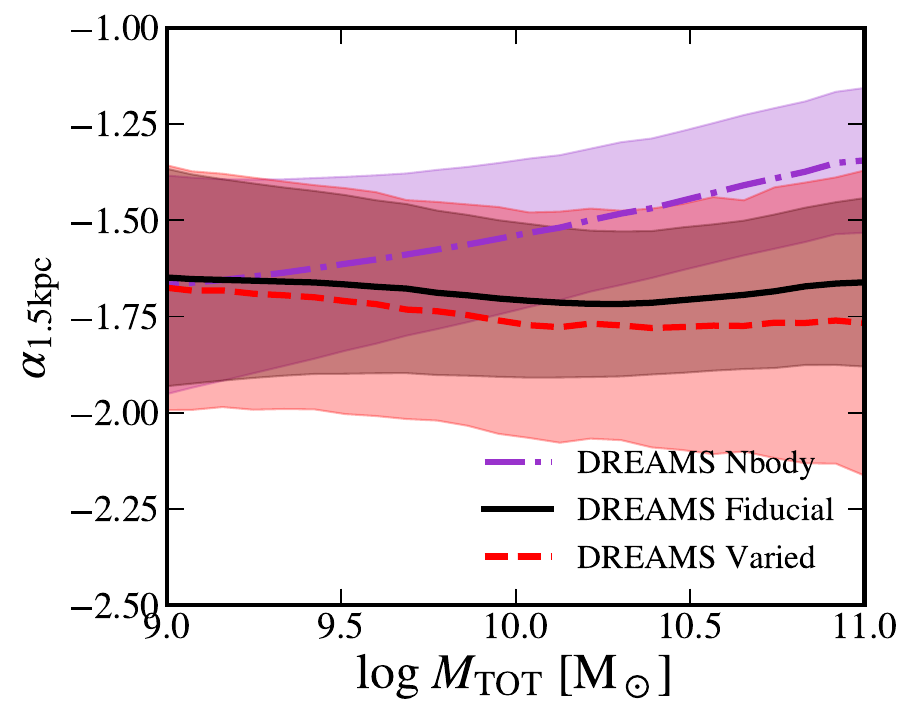}
    \caption{\textit{Left:} DM density profiles for three randomly selected satellites in the DREAMS CDM suite, covering a wide range of satellite mass.
    Black lines show the profiles from the hydrodynamic suite, and the dot-dashed purple lines show the same satellites from the N-body suite.
    The vertical gray band shows the region over which the slope of the density profile is calculated.
    Signs of adiabatic contraction are evident within 2~kpc as the hydrodynamic density profile becomes cuspier than the corresponding N-body profile.
    \textit{Right:} The slope of the DM density profile at 1.5~kpc, $\alphaslope$, versus total satellite mass for the emulated DREAMS Fiducial~(black) and Varied~(red) datasets. 
    For comparison, the results from the DREAMS N-body simulations are also provided in purple. Lines indicate the mean value with the band covering the $1\sigma$ spread.
    The density slope is evaluated between 1.3--1.7~kpc of the satellite center, where effects from gravitational softening are reduced. No stellar mass cut is applied for the satellites in this figure to allow for a fair comparison with the N-body suite. In general, the inclusion of baryons steepens the slope across all halo masses, but the range in $\alphaslope$ appears to be largely driven by halo-to-halo variance and uncertainty in the host halo mass. }
    \label{fig:slope}
\end{figure*}

The blue bands in the left and right panels of Figure~\ref{fig:mass_func} show the 1$\sigma$ region associated with the SAGA data. We caution that the DREAMS and SAGA results should only be compared heuristically at this stage, as a more precise comparison requires obtaining mock SAGA observations from the DREAMS data that better capture the host and satellite selections and also model the interlopers---a study that we leave for future work. A preliminary comparison at this stage is still useful, however, for contextualizing the DREAMS satellite data against observations.

As the left panel of Figure~\ref{fig:mass_func} shows, the SAGA SMF matches the emulated DREAMS Fiducial and Varied datasets for satellites with $\mstar > 10^{7.5}~\mdot$. The SAGA galaxies have an average of $0.38^{+0.97}_{-0.27}$ satellites with $\mstar>10^9~\mdot$ and $1.89^{+4.13}_{-1.08}$ with $\mstar<10^9~\mdot$, consistent with both emulated datasets.
The right panel of Figure~\ref{fig:mass_func} also shows a good correspondence in the satellite radial distribution between the SAGA and emulated DREAMS datasets, especially at large projected radii. However, at small radii, $R_{\rm sat}\lesssim 25$~kpc, more satellites are observed in DREAMS than in SAGA.

\section{Individual Satellite Properties}
\label{sec:individual}

This section characterizes the internal properties of the satellites in the DREAMS CDM suites.
Section~\ref{sec:dm} focuses on the DM distribution for satellites and Section~\ref{sec:stellar} studies the stellar properties for the subset of systems with a stellar mass $\mstar > 10^{7.5}~\mdot$. 
Satellites with a stellar mass close to $10^{7.5}~\mdot$ only contain $\sim$$100$ star particles.
While this sufficiently resolves the mass of the halo, it may not resolve a satellite's inner structure. 
Appendix~\ref{app:resolution} provides an expanded discussion of the resolution effects that impact these satellite properties.

\subsection{Dark Matter Profiles}
\label{sec:dm}

The inner DM density profile of satellite galaxies is a primary diagnostic for understanding the interplay between CDM structure formation and baryonic physics~\citep{2014DiCintio, 2015Chan, 2016Tollet, 2020MNRAS.497.2393L, Anbajagane2022baryonic, Azartash2024}.
Baryonic feedback can shape the distribution of DM around satellites, especially in systems that experience episodic bursts of star formation~\citep{2010Governato,Pontzen2012,Zolotov2012, Brooks2014, 2020MNRAS.497.2393L,2024Mostow}.
These bursts cause rapid changes to the central potential, which can reduce the inner DM density.
In the TNG simulations, satellites do not undergo successive bursts of star formation~\citep{2018Lovell,2018Pillepicha}.
Instead, the TNG galaxies undergo adiabatic contraction: the condensation of baryons into the satellite's center deepens the potential, pulling in DM and steepening the density profile~\citep[][Garcia~et~al.~in~prep]{1986Blumenthal,2004Gnedin,2018Lovell,2023Velmani}.

The left panel of Figure~\ref{fig:slope} provides the DM density profiles for three randomly selected satellites in the DREAMS hydrodynamic suite~(black), along with the profiles for their counterparts in the N-body suite~(dot-dashed purple).\footnote{To limit resolution effects, we only consider satellites with $\mtot>10^9~\mdot$. As detailed in Appendix~\ref{app:resolution}, resolution tests show that  $\alphaslope$ is robust for satellites with $\mtot>10^{9.5}~\mdot$ but has some artificial flattening between $10^9~<~\mtot<10^{9.5}~\mdot$.}
The three examples span over two-orders-of-magnitude in satellite total mass.  In all three cases, the profile from the hydrodynamic run is cuspier than its N-body counterpart within 2~kpc due to adiabatic contraction.
The effect of the adiabatic contraction increases with satellite mass, causing more significant deviations from the N-body partner.

The vertical gray band in Figure~\ref{fig:slope}~(left) indicates the region over which the DM density slope, $\alphaslope$, is evaluated.
As described in Section~\ref{sec:calcs}, the slope is calculated at 1.5~kpc using a finite difference between 1.3 and 1.7~kpc from the satellite's center.
We ensure that the minimum of 1.3~kpc is beyond the radial limit, approximated as $2.8\times\epsilon_{\rm grav}$~\citep{2003Power}, where results may be affected by the gravitational softening ($\epsilon_{\rm grav} = 0.441$~kpc).
We choose a range of 0.4~kpc between the minimum and maximum radii of the finite difference, but we find that the results are not sensitive to this choice.  

The purple dot-dashed line in the right panel of Figure~\ref{fig:slope} shows the mean inner slope for the N-body simulations, with the shaded band indicating the 1$\sigma$ spread. 
The mean slope exhibits a clear upward trend, increasing from  $\alphaslope\sim-1.6$ to $-1.35$ moving from $\mtot=10^9$ to $10^{11}~\mdot$.
This is a natural consequence of DM density profiles that resemble the Navarro-Frenk-White~(NFW) profile~\citep{1997ApJ...490..493N}.
The NFW profile transitions from $\alpha\sim-1$ in the inner region  to $\alpha<-2$ at larger radii, and the point of that transition (e.g., the scale radius) increases with $\mtot$.
As a result, for high-mass galaxies, $\alphaslope$ probes the density profile at radii that are comparatively closer to the center.  This explains the trend of steeper slopes for less-massive satellites for the N-body data.

While the trend of $\alphaslope$ for the satellites in the N-body simulation is not constant with $\mtot$, it still acts as a useful comparison to understand how the inclusion of baryons and parameter variations affects the DM halo.
To this end, Figure~\ref{fig:slope}~(right) compares the N-body results with the emulated DREAMS Varied dataset~(red), which include the effects of baryons. The DREAMS Varied results deviate from the N-body results for satellites with $\mtot\gtrsim10^{9.5}~\mdot$, indicating the mass range where adiabatic contraction starts to affect the DM density at 1.5~kpc.

Consistent with previous findings for the TNG galaxy-formation model~\citep[][Garcia~et~al.~in~prep]{2018Lovell,2023Velmani}, no satellites at any mass scale resolved here have cored density profiles with $\alphaslope \gtrsim-1$.
For the DREAMS Varied dataset, the average value for the slope is $-1.75$ across all satellite masses, except for a slight flattening for $\mtot\lesssim 10^{10}~\mdot$. This result is also in agreement with previous studies of group- and cluster-sized halos, with $\mtot\ge 10^{13}~\mdot$, in the TNG simulations, which likewise find that satellites maintain cuspy inner density profiles across all resolved mass scales \citep{Farahi2022profile}.

The black line and band in Figure~\ref{fig:slope}~(right) shows the results for the emulated DREAMS Fiducial dataset.  This dataset does not include physics variations, isolating the effects from intrinsic halo-to-halo variance and uncertainty on the host halo mass for a fixed set of astrophysical and cosmological parameters. Comparing the black and red bands gives a sense of the relative impact of the model variations compared to the intrinsic halo-to-halo variance and uncertainty on the host halo mass.  The effect is minimal, with the average value for the slope remaining approximately $-1.67$ across the full mass range.

The mean DM density slope, $\alphaslope$, is not significantly affected once baryonic physics is varied; however, there is an increase to the 1$\sigma$ spread about the mean due to variations in the energy of SN winds, \ew. 
Across all resolved satellites, increasing \ew\,results in a higher DM density at a fixed stellar mass.  This results from the fact that increasing \ew\,reduces the galaxy's stellar mass.
The 1$\sigma$ spread increases by $\sim$$10\%$ toward flatter profiles and $\sim$$70\%$ toward cuspier profiles for satellites with $\mtot>10^{10}~\mdot$.
For lower-mass satellites, there is only a slight, $\sim$$~20\%$, greater spread toward cuspier profiles.

We also examine whether $\alphaslope$ varies with the mass of the host galaxy.
We find no measurable dependence of $\alphaslope$ on the host total mass within the range $(0.5-2.0)\times10^{12}~\mathrm{M_\odot}$, suggesting that host halo potential has no impact on the inner halo profiles at this mass range.
Therefore, within the TNG model, uncertainty in the model parameters can lead to a wider range of DM slopes than the fiducial case, but uncertainty in the host halo mass does not contribute strongly for the MW.

\subsection{Stellar Properties}
\label{sec:stellar}

\begin{figure*}
    \centering
    \includegraphics[width=\columnwidth]{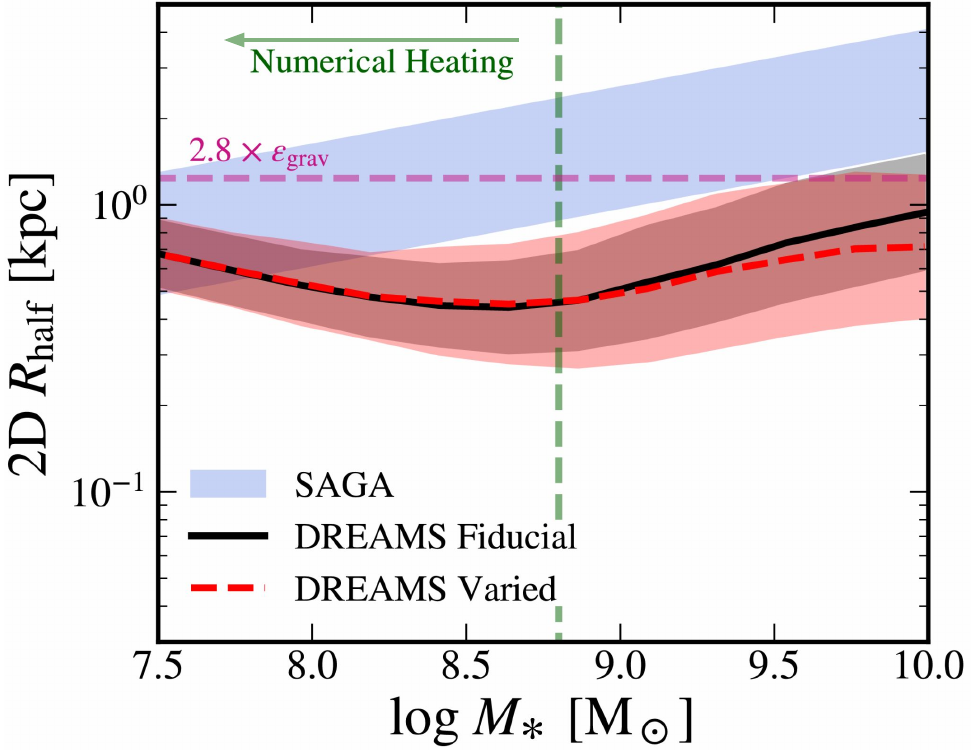}
    \includegraphics[width=\columnwidth]{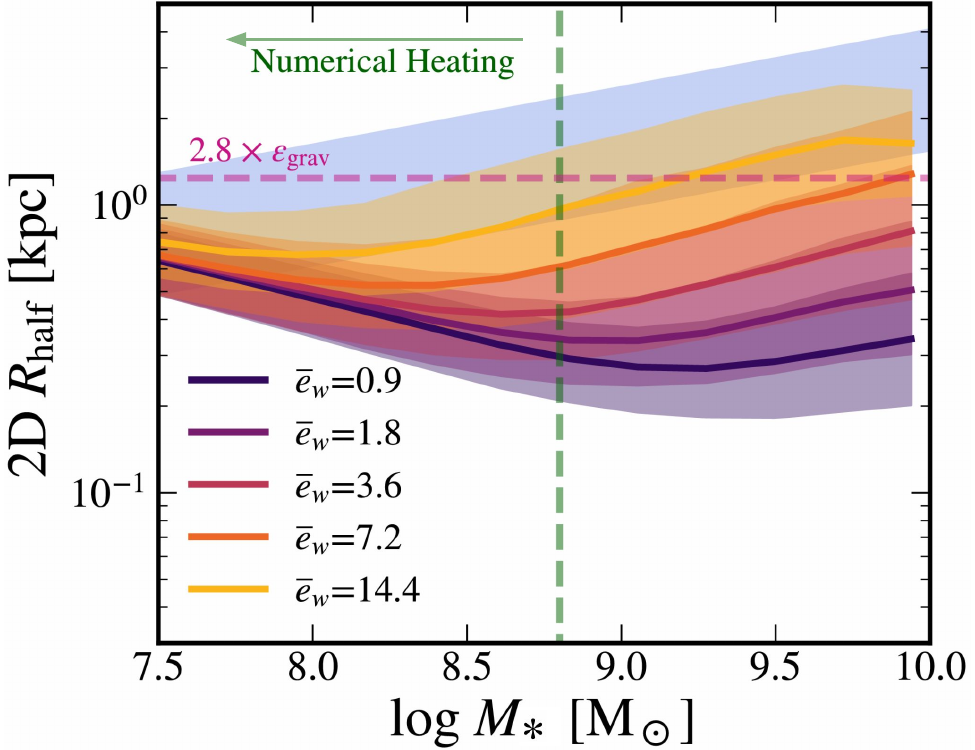}
    \caption{\textit{Left:} Satellite size-mass relation for the emulated DREAMS Fiducial~(black) and Varied~(red) datasets.  The observed half-light radii from SAGA ($1\sigma$ spread) are shown by the blue band~\citep{SAGAVI}.
    Size is measured using the r-band 2D circularized stellar half-light radius for all datasets, see Section~\ref{sec:calcs} for details. Satellites falling below the pink-dashed line or to the left of the green dashed line are subject to numerical resolution or heating effects.  The DREAMS simulations do not produce the SAGA results in the top-right of the panel, where numerical effects should be under control.
    \textit{Right:} The size-mass relation is shown for different values of the wind energy \ew, the dominant source of physical uncertainty in the model.  The varied astrophysical and cosmological parameters in the DREAMS model are not sufficient to bring the results into correspondence with SAGA.  For both the left and right panels, the mean size-mass relation is indicated by the colored lines, while the bands indicate the $1\sigma$ spread.  
    }
    \label{fig:size}
\end{figure*}

This subsection explores the stellar distribution of the DREAMS satellites, specifically their half-light radii. The left panel of Figure~\ref{fig:size} shows the 2D stellar half-light radius, $R_{\rm half}$, versus the total stellar mass, $M_*$,
from the SAGA Survey as a blue shaded band~\citep{SAGAVI}.
To ensure a consistent comparison with our simulated results, we use the circularized r-band size-mass relation from SAGA rather than the semi-major axis lengths presented in their main results.
Unlike Section~\ref{sec:population}, the SAGA results here include all 101 host galaxies in their sample.
However, SAGA finds that only satellite abundances are strongly correlated with host halo mass \cite[see Appendix~A in][]{SAGAIII}, so we do not expect this to significantly affect our results.

The observations can be compared to the emulated DREAMS Varied~(red) and Fiducial~(black) dataset. However, two significant sources of numerical uncertainty must be considered when interpreting the DREAMS results.  First, with a gravitational softening of $\epsilon_{\rm grav} =0.441$~kpc, it is not possible to accurately resolve satellites with a half-light radius smaller than $2.8$ times this value~\citep{2003Power}.
Similarly, \cite{2020Ludlow} showed that sizes below the convergence radius, $0.05~\times~l$ (where $l$ is the average interparticle spacing) are not accurate.
These metrics give similar lower bounds for the radius, 1.2~kpc and 1.8~kpc, respectively.
The horizontal dashed pink line in Figure~\ref{fig:size} indicates $2.8~\times~\epsilon_{\rm grav}$.

Separately,~\citet{2019Ludlow-equi} showed that cosmological simulations with unequal-mass stellar and DM particles, where $M_{\rm DM}>M_{\rm b}$, can suffer from spurious energy transfer from the more-massive DM particles to less-massive star particles.
This numerical effect, a consequence of mass segregation, can cause the stellar component to artificially heat and expand, leading to larger galaxy sizes over time.
This effect is most pronounced for poorly resolved galaxies and likely explains the increasing sizes with decreasing $\mstar$ seen in the emulated DREAMS datasets.
We place a vertical green dashed line at $\mstar=5\times10^8~\mdot$ to mark where this effect starts to dominate the size measurements.

Even accounting for these numerical effects, the observed size-mass relation for SAGA should be resolved for DREAMS satellites with $\mstar\gtrsim 10^9~\mdot$.  However, as seen in Figure~\ref{fig:size}, this is clearly not the case, with both the emulated DREAMS Varied and Fiducial datasets falling below the blue band in this range.  This discrepancy is not resolved through variations of the astrophysical and cosmological parameters considered here.  Of these, the energy of the SN winds, \ew, dominates the uncertainty on the stellar half-mass radius.  The right panel of Figure~\ref{fig:size} shows how the 2D $R_{\rm half}$ varies with \ew.  
The specific choice of \ew~affects the magnitude of the radius, with higher values creating more extended satellites in the high-mass regime, $\gtrsim 10^9~\mdot$.

Despite this behavior, the left panel of Figure~\ref{fig:size} shows that the average 2D~$R_{\rm half}$ is smaller in the DREAMS Varied suite than in the Fiducial suite for $M_* \gtrsim 10^9~\mdot$.
This effect is tied to the dependence of the stellar mass on \ew. 
As \ew\,increases, the stellar mass for all satellites decreases, making those with $\mstar>10^9~\mdot$ rare, occurring in only 7\% of systems. However, at low \ew, satellites with $\mstar>10^9~\mdot$ are more common, occurring in 93\% of systems.
Therefore, when taking the average size of all satellites with $\mstar>10^9~\mdot$, the average shifts downward because the less massive galaxies are more common. 

Generally, the SAGA satellites exhibit larger half-light radii at fixed stellar mass than the mean DREAMS relation, particularly for $\mstar>10^{9}~\mdot$.  The discrepancy can either be due to (1)~observational realities not accounted for in the simulation data or (2)~incomplete physics within the TNG galaxy-formation model.

Direct comparison between the sizes of simulated galaxies and observational data is complex.
For example, the translation from stellar mass to light is uncertain and can be variable based on gradients in metallically and dust~\citep{2013Conroy}.
Additionally, observational data are subject to surface brightness limits and signal-to-noise cuts, which can lead to difficulties in detecting faint, extended components that are normally included in simulation analyses~\citep{SAGAI,SAGAII}.
Additionally, within the TNG model, galaxy sizes are known to decrease with increasing resolution~\citep{2019Pillepich}---see Appendix~\ref{app:resolution}.
While the method of creating 2D luminosity maps described in Section~\ref{sec:calcs} is a step toward a fair comparison between simulation and data, we do not create mock SAGA images, and these intrinsic and numerical challenges in comparing size metrics should be kept in mind. 
That being said, the fact that the DREAMS SMF and radial distributions are similar to the SAGA results\footnote{See Figure~\ref{fig:mass_func}.  Note that the distributions do not significantly change with the inclusion of the additional SAGA hosts.} suggests that the size-mass discrepancy is likely not dominated by limitations in observed surface-brightness limits or incompleteness. 
Additionally, we explore the impact of the host halo mass on the satellite size-mass relation and find that the stellar half-light radii show no significant variation with the total mass of the host galaxy.
Instead, the discrepancy may arise from missing physics within the TNG model.  

High-resolution zoom simulations that resolve a more dynamic, multi-phase interstellar medium~(ISM) and feature more `bursty' star formation can produce satellites with a wider range of sizes~\citep{2017Fitts, Applebaum2021}.  Such models tend to push stars in dwarfs to larger radii via fluctuations in the gravitational potential~\citep{Graus2019, Riggs2024}.  The lack of such fluctuations in the TNG galaxy-formation model may result in smaller galaxy sizes.
The TNG model's effective equation of state for its ISM~\citep{2003Springel}, as well as its more uniform feedback from star formation and mass loading of the winds, may prevent the extreme variations in stellar density that are needed to match the observed sizes of these satellites~\citep{2024Shen-size}.

\section{Discussion and Conclusions}
\label{sec:discussion}

This paper utilized a new suite of 1,024 CDM galaxies from the DREAMS Project to quantify the dominant sources of uncertainty in modeling satellites about MW-mass hosts.  The analysis focused on properties of the satellite population~(SMF and radial distribution), as well as structural properties of individual satellites (DM density and stellar half-light radius). Each DREAMS MW-mass host is generated for a unique initial condition and set of five varied parameters from the TNG model.  The latter includes the SN wind energy, \ew, and speed, \kw, as well as the AGN feedback parameter, \agn.  The varied cosmological parameters are \om\, and $\sigma_8$.

To effectively interpolate over the discrete parameter sampling, we emulated datasets that take as input the varied simulation parameters and generate satellite populations around MW-mass hosts.  Using these datasets, the key findings are:
\begin{itemize}
    \item Halo-to-halo variance, arising from different formation histories and environments, is the dominant source of uncertainty in most satellite properties. For the SMF, radial distribution, and 2D half-light radius, the halo-to-halo scatter outweighs uncertainties introduced by varying the SN and AGN feedback parameters in the TNG galaxy-formation model. The DM density slope, however, is an exception. While the mean slope is robust to model variations, the 1$\sigma$ spread depends on the astrophysical parameters, mainly the supernova wind energy. The varied cosmological parameters do not have a measurable effect on any of the satellite properties.
    \item The satellite SMF and radial distribution largely match the 1$\sigma$ spread for the corresponding SAGA distributions. The one exception is that there are less satellites within $R_{\rm sat} \lesssim 25$~kpc in SAGA than in the DREAMS simulations. 
    \item At mass scales $M_{\rm TOT} > 10^9~\mdot$, the DREAMS satellites are not significantly cored, even when accounting for variations in the TNG feedback parameters.  Instead, adiabatic contraction plays an important role in shaping the inner slope of their DM profiles.  
    \item At the well-resolved high-mass end ($\mstar>10^9~\mdot$), the DREAMS satellites are systematically smaller than those observed in the SAGA survey.  This discrepancy is not solved with the parameter variations considered.   
\end{itemize}

While uncertainties from baryonic physics are subdominant, when they do have an effect, it is driven almost entirely by the SN wind energy, \ew. For example, increases to \ew\,tend to decrease the stellar mass of a satellite and increase its stellar half-light radius, especially for systems with $\mstar\gtrsim10^9~\mdot$. 
The SN wind speed, \kw, has a minor secondary effect. 
Likely, the implementation of the SN wind speeds in TNG reduces the effect of \kw\,on lower-mass satellites because of the minimum speed floor of 350~km/s (see Equation~\ref{eq:speed}). 
At $z=0$, satellites with a local DM velocity dispersion less than 50~km/s will have a wind speed of 350~km/s, assuming the fiducial TNG value of \kw.  For the DREAMS satellites, this means that satellites with $\mstar\lesssim3\times10^{8}~\mdot$ will typically be assigned the minimum wind speed.

Additionally, variations to the AGN feedback parameter, \agn, have no measurable impact on the satellite properties.  In the TNG model, for a satellite to be seeded with a BH, it must have a halo mass above $7.2\times~10^{10}~\mdot$ before it is accreted onto the host galaxy. Because only the most massive satellites have a BH, \agn\,has little overall effect on the general satellite properties.
While the central BH in the host MW galaxy could potentially affect the satellites by expelling surrounding gas, this does not appear to be a significant factor for the satellite properties investigated here.

Even with the variations in these baryonic physics parameters, discrepancies persist between the DREAMS and SAGA size-mass relation.  While numerical resolution and heating effects can complicate this comparison, the discrepancy is still present for the robust high-mass satellites ($\mstar>10^9~\mdot$).
It is possible that other variations within the TNG model, such as the SN wind velocity floor, $v_{w,\mathrm{min}}$, or the initial mass function~(IMF), may affect the size of the satellites and could increase the satellite sizes while keeping the other properties consistent with observations.
However, a more likely scenario is that this discrepancy is due to missing physics that is required to create more diverse stellar distributions.
For example, other galaxy-formation models, such as FIRE~\citep{FIRE1, FIRE2}, NIHAO~\citep{Wang2015}, or Marvel/Justice League~\citep{Applebaum2021} resolve local star-forming regions of cold gas that can affect the internal properties of satellites around MW-mass galaxies, particularly by causing fluctuations in the gravitational potential well that may also lead to bigger satellites~\citep{Brooks2014, Graus2019, Riggs2024}, or at least prevent adiabatic contraction.
Updating the TNG model to explicitly model the multiphase ISM and SN feedback, rather than adopting the effective model from \cite{2003Springel}, and further tuning additional simulation parameters using the method in Rose~et~al.~(submitted), may bring the internal satellite properties into better agreement with observations.  

One source of uncertainty on our results is the halo mass of the MW, which is explicitly modeled in the DREAMS simulations to fall within the observational uncertainty~\citep{1999Wilkinson, 2010Watkins, 2014Kafle, 2018Sohn, 2019Eadie, 2022Wang}.
As detailed in Sections~\ref{sec:population}~and~\ref{sec:individual}, we binned the satellites in three groups based on their host's total mass, similar to the right panel of Figure~\ref{fig:size}.
We find that the host's total mass only affects the SMF in the DREAMS simulations, resulting in $2-3\times$ more satellites for a system with $\mtot\sim2\times10^{12}~\mdot$ than a system with $\mtot\sim5\times10^{11}~\mdot$.
The other three satellite properties show a minimal and subdominant dependence on the host's total mass.

A main conclusion of this work is that halo-to-halo variance, which can arise from the environment around the host and its specific accretion history, is a dominant source of uncertainty in modeling populations of satellites as well as their internal properties.  Importantly, this variance dominates over uncertainties in the baryonic physics modeling within TNG.  There are two critical next steps to this Project to further our understanding.  The first is to remove the isolation criterion on the simulated MW-mass hosts and instead focus on systems with e.g., an M31 partner or an LMC satellite.  Placing further restrictive priors on the environment of the host and its accretion history will demonstrate whether the halo-to-halo variance can be reduced when making predictions for our Galaxy.  The other critical step is to repeat these studies with different galaxy-formation models, which will enable the quantification of theoretical uncertainties \emph{between} sub-grid physics assumptions.  This paper demonstrates that these extensions are feasible and establishes a program of study to comprehensively assess theoretical uncertainties in simulations of MW-mass systems that will better inform comparisons to observed satellite populations.

\section*{Acknowledgements}

We thank the Simons Foundation for their support in hosting and organizing workshops on the DREAMS Project. 
Additionally, we gratefully acknowledge the use of computational resources and support provided by the Scientific Computing Core at the Flatiron Institute, a division of the Simons Foundation.
The authors acknowledge Research Computing at The University of Virginia for providing computational resources and technical support that have contributed to the results reported within this publication (URL: \url{https://rc.virginia.edu}), as well as Princeton University's Research Computing resources.
The authors also thank the SAGA team for providing the circularized satellite measurements used for comparison in this work.
ML is supported by the Simons Investigator Award. 
XS acknowledges the support of the NASA theory grant JWST-AR-04814. KEK is supported by the National Science Foundation Graduate Research Fellowship Program under Grant No.~DGE-2444107. 
AMG, NK, PT, and AF acknowledge support from the National Science Foundation under Cooperative Agreement 2421782 and the Simons Foundation grant MPS-AI-00010515 awarded to NSF-Simons AI Institute for Cosmic Origins (CosmicAI, \href{https://www.cosmicai.org/}{https://www.cosmicai.org/}).

\section*{Data Availability}

The data and code used to produce this paper can be made available upon reasonable request to the corresponding author.
Further information and source code for NeHOD can be found at \url{https://github.com/trivnguyen/nehod_torch}.

\clearpage
\appendix
\section{Emulator Training and Validation}
\label{app:nehod}

\setcounter{equation}{0}
\setcounter{figure}{0} 
\setcounter{table}{0}
\renewcommand{\theequation}{A\arabic{equation}}
\renewcommand{\thefigure}{A\arabic{figure}}
\renewcommand{\thetable}{A\arabic{table}}

In this work, we leverage the NeHOD emulator~\citep{NeHOD} to efficiently explore the 5D simulation parameter space within the DREAMS suite. As described in Section~\ref{sec:nehod}, generating satellite populations for larger numbers of parameter combinations is computationally prohibitive with hydrodynamic simulations.
Utilizing the emulator allows us to overcome this limitation by rapidly producing statistically accurate satellite populations for any set of input parameters.

NeHOD is a two-step hierarchical generative model.
The first step models the properties of the host halo, which are subsequently used to condition the second step, which generates the full satellite galaxy population.
The training procedure is described here, along with various validation tests, to understand the extent to which the emulator is replicating the properties of the DREAMS simulations.

The first stage consists of a Neural Posterior Estimator (NPE) implemented as a normalizing flow.
This model is trained to learn the posterior distribution of three halo properties: the host stellar mass, the host halo mass, and the number of satellites with $\mstar>10^{6.5}~\mdot$.
We slightly extend the satellite mass cut below the resolution limit used in this analysis, $\mstar=10^7~\mdot$, so that NeHOD does not learn the hard cut at this limit.
This distribution is conditioned on the five simulation parameters (\om, \s8, \ew, \kw, \agn).
The normalizing flow architecture consists of six neural spline flow layers with four hidden layers.

We split the DREAMS dataset into training and validation datasets, with 921 and 103 simulations, respectively.
The model is trained for a maximum of $10^4$ steps using a batch size of 32 and optimized with the AdamW optimizer~\citep{adamW}.
We performed a 50-trial hyperparameter search using \textsc{optuna}~\citep{optuna} to optimize the learning rate, weight decay, dropout rate, and number of flow transforms.
The best performing trial achieved a final validation loss of $2.09$, utilizing a learning rate of $2.5\times10^{-2}$, a weight decay of $6.6\times10^{-3}$, a dropout rate of $7.6\times10^{-4}$, and a single flow transform.

\begin{figure*}
    \centering
    \includegraphics[width=\textwidth]{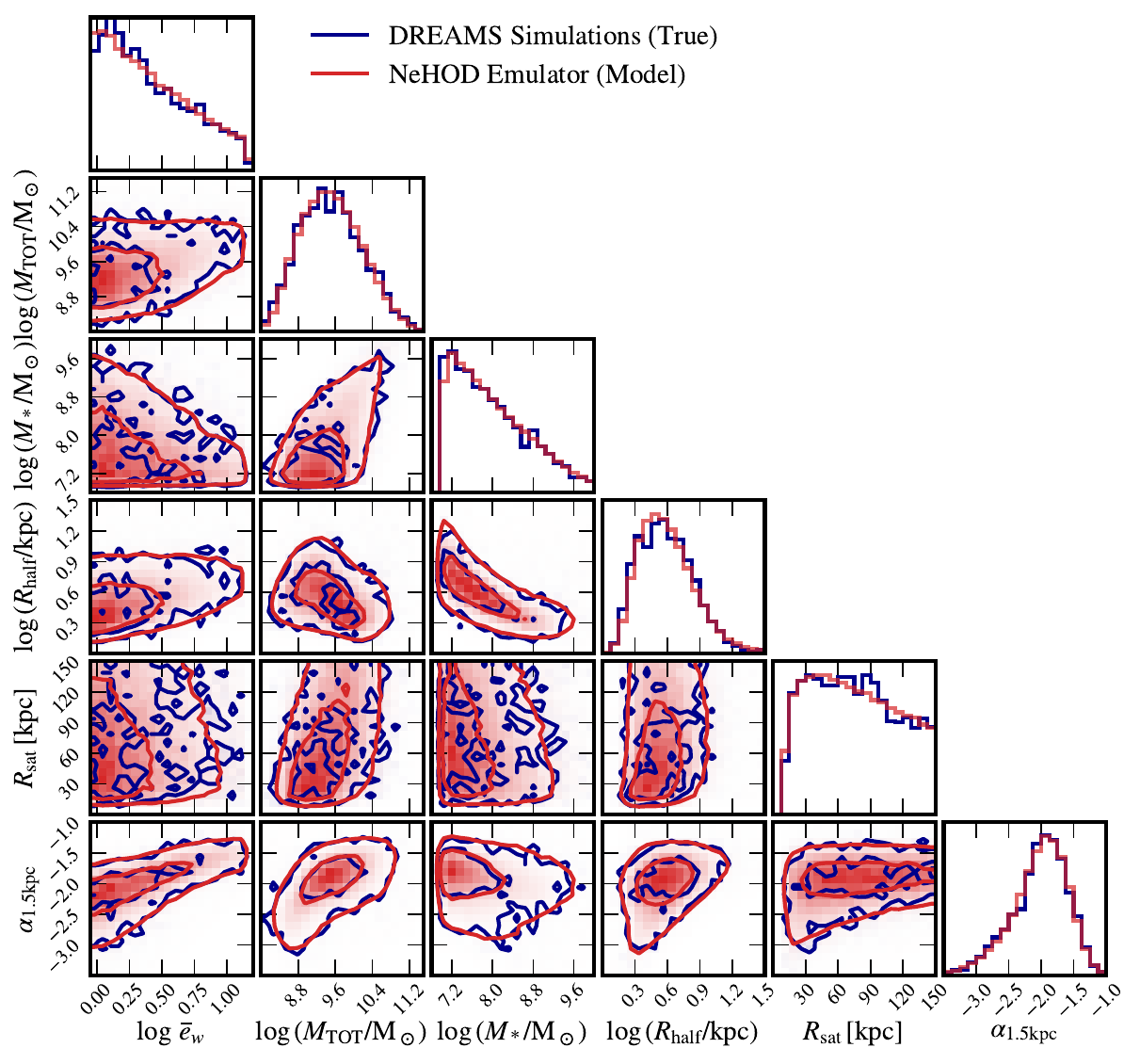}
    \caption{A validation corner plot comparing the distributions of five key satellite properties from the original DREAMS simulations~(blue) and those generated with the emulator~(red; shown as `DREAMS Varied' in the main text).
    The diagonal panels show the 1D probability distributions and the off-diagonal panels show the 2D joint distributions.
    The agreement between the simulations and the emulator demonstrates that NeHOD reproduces not only the distribution of individual properties, but also the complex correlations between them.}
    \label{fig:nehod_corner}
\end{figure*}

The second step of the emulator is a Variational Diffusion Model~\citep[VDM;][]{2021Vahdat,2021Song,2023Kingma}, which generates the 7D vector of orbital and physical properties for each satellite. Specifically, the model treats the satellites as a point cloud where each point has the attributes of the given satellite (e.g., position, velocity, mass).
This approach avoids the need for voxelation, which can lose spatial information.  The VDM generates satellite properties by utilizing a transformer-based noise model, which processes the entire satellite population at once.  It is conditioned on an 8D  vector comprising the five simulation parameters and the three halo properties generated by the normalizing flow in the first step.
The model is designed to handle variable satellite populations up to a maximum of 50 satellites per host through the use of an attention mask.
We find that only 10 samples out of the $10^5$ used in the analysis presented here reach the maximum number of 50 satellites, indicating that this limit is not affecting our results.
This model is trained for $5\times10^4$ steps with a batch size of 32 and the AdamW optimizer.
A separate 50-trial hyperparameter optimization is conducted to optimize the VDM learning rate, weight decay, and the noise schedule limits, $\gamma_\mathrm{min}$ and $\gamma_\mathrm{max}$.
The best-performing VDM trial achieved a validation loss of $-6.7$ utilizing a learning rate of $5.3\times10^{-2}$, a weight decay of $2.8\times10^{-4}$, a $\gamma_\mathrm{max}$ of 14, and a $\gamma_\mathrm{min}$ of $-13$.
To incorporate the epistemic uncertainty from the emulator training, we generate datasets using an ensemble of the five best-trained normalizing flows and VDMs, each producing 20\% of the final dataset.

\begin{figure*}
    \centering
    \includegraphics[width=0.49\textwidth]{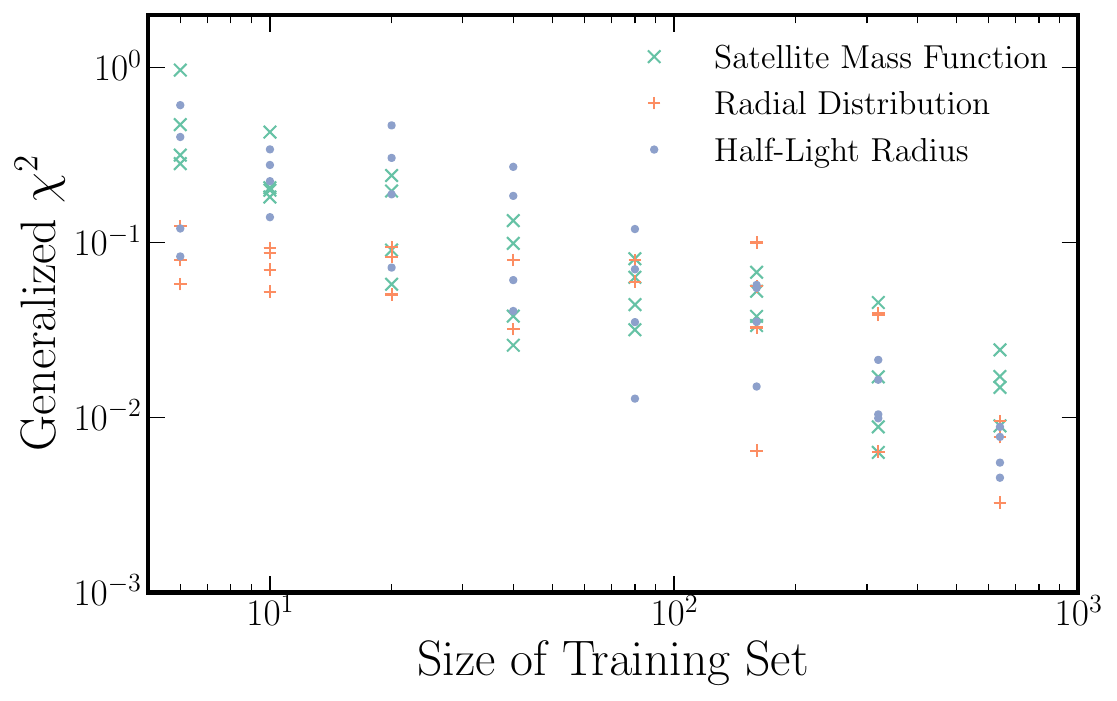}
    \includegraphics[width=0.49\textwidth]{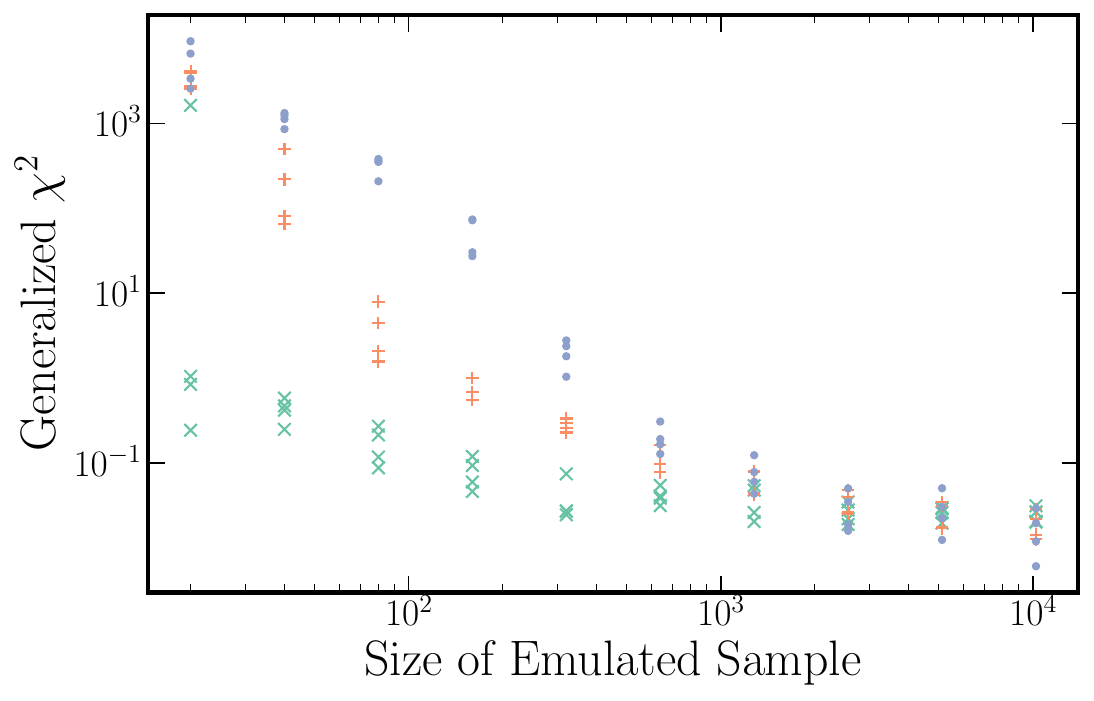}
    \caption{Emulator performance, quantified by the generalized $\chi^2$ statistic (see Equation~\ref{eq:chisquared}), for three satellite properties: the satellite mass function~(green x), the radial distribution~(orange +), and the half-light radius~(blue dot).  Each test is repeated four times to capture the scatter from the training process. 
    A lower $\chi^2$ indicates a better agreement between the emulated data and the original DREAMS simulations.  
    \textit{Left:}~The generalized $\chi^2$ as a function of the number of simulations used to train the NeHOD emulator.
    The $\chi^2$ value and its scatter decrease with the size of the training set, but do not converge to a specific value.
    While the number of training simulations do not converge, the generalized $\chi^2$ value at 640 simulations is already quite small; see Figure~\ref{fig:traning_validation} for a visual representation.
    \textit{Right}: The generalized $\chi^2$ as a function of the number of emulated samples generated by the trained emulator used in the main text.
    The $\chi^2$  value and spread decrease as the emulated sample grows, converging after $\sim$$2\times 10^3$ samples are generated.
    }
    \label{fig:training_size}
\end{figure*}

To ensure that the emulated data is a good reproduction of the simulations, we validate the model by comparing the statistical distributions of satellite properties.
Figure~\ref{fig:nehod_corner} shows a corner plot comparing the 1D and 2D distributions of one input parameter, \ew, and five satellite properties from the DREAMS simulations~(blue) to those generated by NeHOD~(red; shown as `DREAMS Varied' in the main text).
The contours show the $1\sigma$ and $2\sigma$ containment regions.
The close agreements between the contours and histograms confirm that NeHOD can successfully capture the complex correlations between different satellite properties and the input parameters.
We do not show the full corner plot of five input parameters and seven output parameters for readability, but confirm that all parameter combinations show good agreement between the simulations and emulated samples.

Next, we investigate whether the size of the DREAMS simulation suite, as well as the number of emulated samples, is sufficient to fully represent the population of Milky Way~(MW)-mass galaxies in the DREAMS suite.
To test this, we train the NeHOD emulator multiple times on increasingly larger subsets of the full simulation suite.
We then generate large mock datasets from each of these emulators and compare their properties to those from the emulator trained on the full dataset.  To quantify this comparison, we use a generalized $\chi^2$ statistic, defined as
\begin{equation}
    \chi^2 = (\bar{y}_1 - \bar{y}_2)^T \cdot C^{-1}_{\mu,\mathrm{total}} \cdot (\bar{y}_1 - \bar{y}_2)  \, ,
    \label{eq:chisquared}
\end{equation}
where $\bar{y}_1$ and $\bar{y}_2$ are the mean vectors of summary statistics from the two emulators (trained on the smaller subset and the full sample, respectively), and $C_{\mu,\mathrm{total}}$ is the total covariance matrix of the difference between these two means, given by $C_{\mu,\mathrm{total}}=C_{\mu,1}+C_{\mu,2}$. 
Here, $C_{\mu,i}$ is the covariance of the mean vector and $i=1,2$ refers to the ground truth `DREAMS Varied' dataset and the smaller emulated dataset.   
Note that $C_{\mu,i}=C_{i}/N_i$,where $N_i$ is the number of galaxies in the emulated dataset used to compute the mean, and $C_{i}$ is the covariance matrix for the dataset given by
\begin{equation}
    C_{i} = \frac{1}{N_i - 1} \sum_{j=1}^{N_i} (y_{i, j} - \bar{y}_i)(y_{i, j} - \bar{y}_i)^T .
\end{equation}


Figure~\ref{fig:training_size} shows the convergence of this generalized $\chi^2$ as the number of training simulations increases (left) and the number of emulated samples increases (right).
A lower $\chi^2$ indicates a better agreement between the emulated data and the original DREAMS simulations.

The emulator's performance is tested for the case of the satellite mass function~(green x), the radial distribution (orange +), and the half-light radius (blue dot).  In each case, we repeat the training and data generation process four times to understand the inherent scatter in the training process---all of which are shown in the Figure~\ref{fig:training_size} panels.  Some datapoints overlap, resulting in fewer than four points for a particular dataset in some columns.

We find that the $\chi^2$ for the best and worst trained emulators continues to decrease with increasing number of simulations. For the case of the emulated samples, the $\chi^2$ converges after $\sim$$2\times10^3$ samples, far below the $10^5$ samples used for the emulated datasets in this paper.

\begin{figure*}
    \centering
    \includegraphics[width=\textwidth]{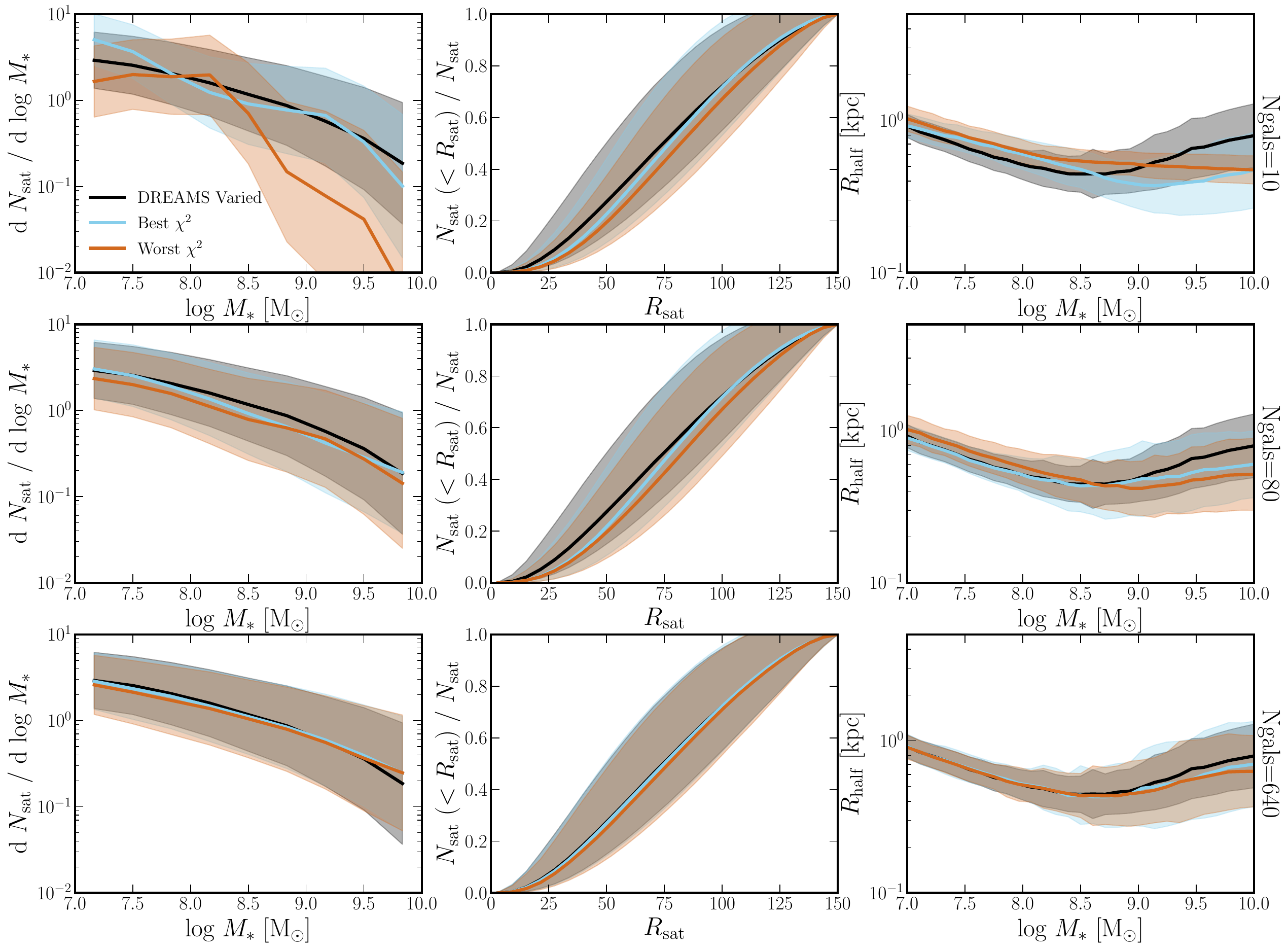}
    \caption{This figure provides a visual representation of the convergence test shown in the left panel of Figure~\ref{fig:training_size}, illustrating what the quantitative $\chi^2$ values correspond to.
    Each row shows the results for a different number of simulations in the NeHOD training set, either 10~(top), 80~(middle), or 640~(bottom).
    The columns correspond to the satellite mass function~(left), radial distribution~(middle), and half-light radius~(right).
    All properties follow the calculations described in Section~\ref{sec:calcs} and show the mean as a solid line and the 1$\sigma$ spread as a shaded band.
    In each panel, the gray band represents the ground truth from the full simulation dataset and is compared against the convergence test that produced the `Best' fit (lowest $\chi^2$; blue) and the `Worst' fit (highest $\chi^2$; orange).
    As the number of simulations increases, the `Worst' fit rapidly converges toward the `Best' fit and ground truth.
    This confirms that a training set of $\sim$$640$ simulations is sufficient to reliably reproduce satellite properties where the worst-fit models are in good agreement with the full simulation data.
    }
    \label{fig:traning_validation}
\end{figure*}

To put the different $\chi^2$ values in context, Figure~\ref{fig:traning_validation} compares the best~(lowest $\chi^2$) and worst~(highest $\chi^2$) trained models for three training-set sizes: 10, 80, and 640 simulations.
While the $\chi^2$ continues to decrease with the number of simulations in the training sample, its values are quite low.  In practice, the worst-trained model with 640 simulations is not significantly different from the DREAMS Varied sample.
For just 80 simulations in the training sample, the means are off by at most 50\%, which only occurs in the highest mass bin of the satellite half-light radii.
We thus conclude that the number of simulations in the DREAMS suite (1,024) is large enough to provide a stable representation of these satellite properties.
Additionally, reducing the total number of simulations to $\sim$512 would not significantly affect the results presented here.

\section{Resolution Tests}
\label{app:resolution}

\setcounter{equation}{0}
\setcounter{figure}{0} 
\setcounter{table}{0}
\renewcommand{\theequation}{B\arabic{equation}}
\renewcommand{\thefigure}{B\arabic{figure}}
\renewcommand{\thetable}{B\arabic{table}}

\begin{figure*}
    \centering
    \includegraphics[width=0.49\textwidth]{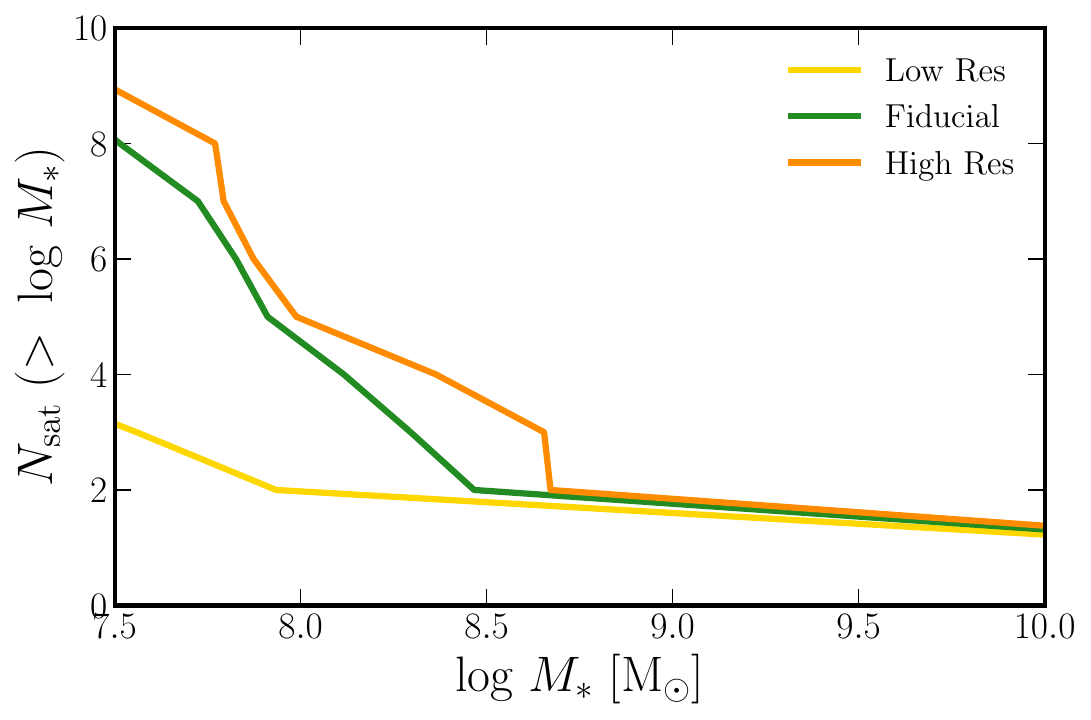}
    \includegraphics[width=0.49\textwidth]{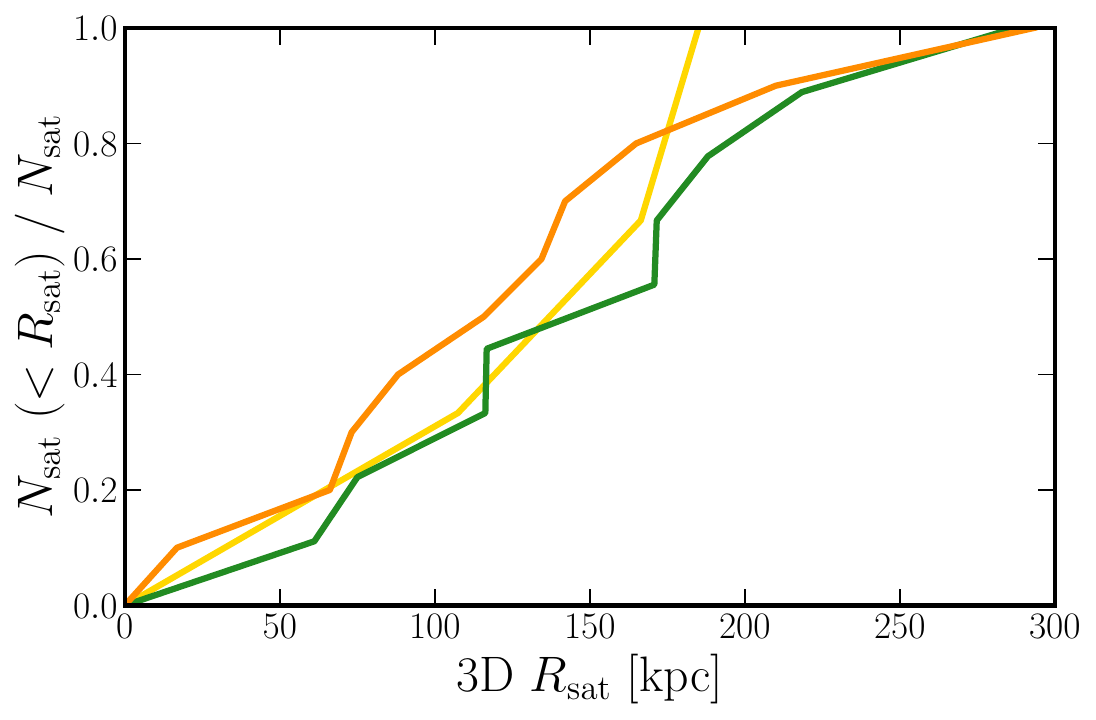}
    \includegraphics[width=0.49\textwidth]{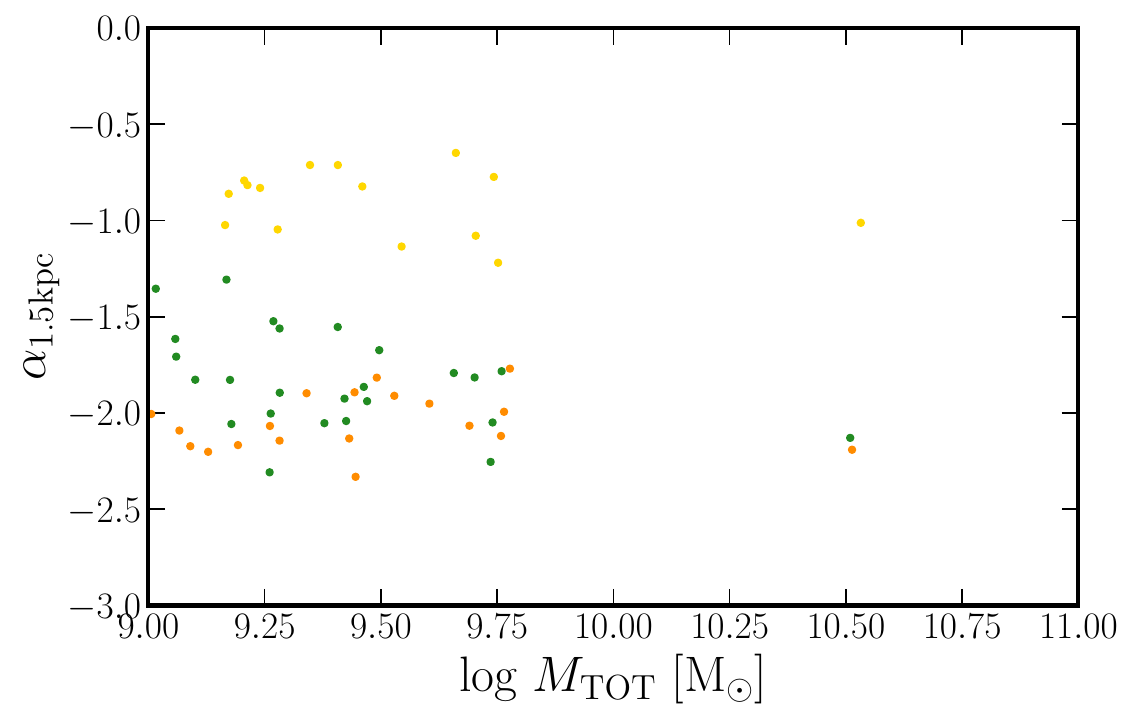}
    \includegraphics[width=0.49\textwidth]{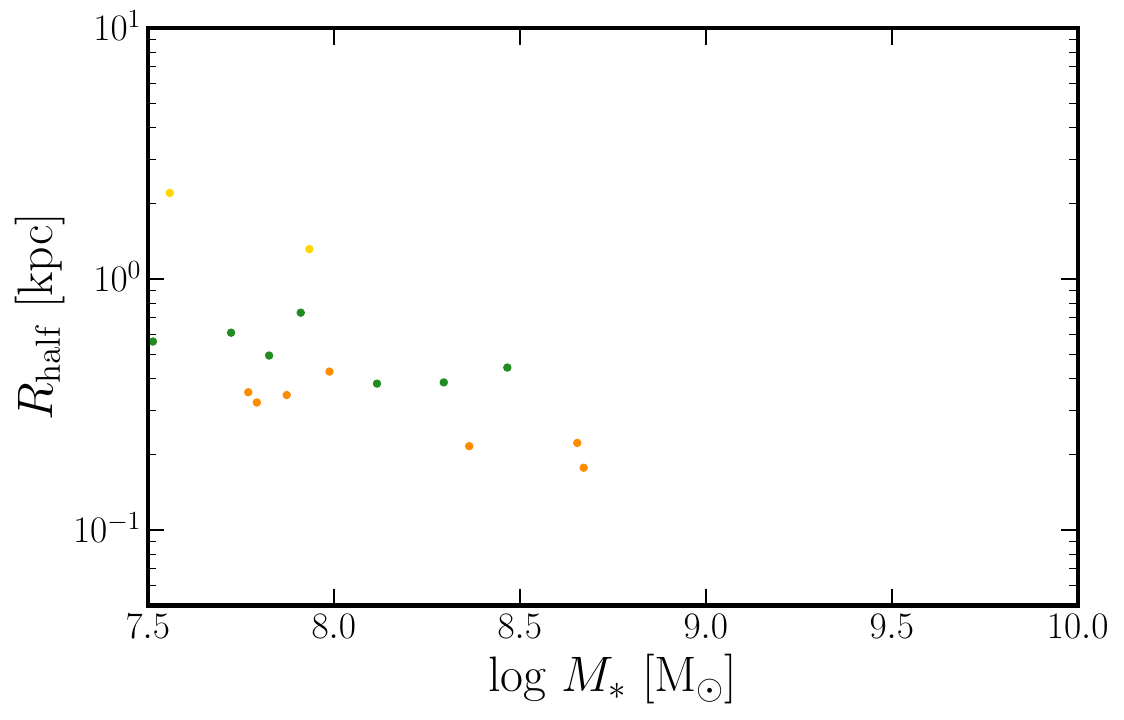}
    \caption{A comparison of satellite properties for a single MW-mass halo taken from the DREAMS CDM suite, simulated at three different resolutions: $8\times$ lower than fiducial (`Low Res', gold), fiducial DREAMS resolution of $M_{\rm DM}=1.8\times10^6~\mdot$ and $M_{\rm b}=2.8\times10^5~\mdot$ (`Fiducial', green), and $8\times$ higher resolution (`High Res', orange).
    \textit{Top-left}: Cumulative SMF; the fiducial run is reasonably converged but $\mstar$ increases with resolution.
    \textit{Top-right}: Normalized cumulative radial distribution; no apparent systematic trend for this galaxy.
    \textit{Bottom-left}: DM density slope at 1.5~kpc; the Fiducial and High-Res simulations converge for $\mtot>10^{9.5}~\mdot$ for this galaxy.
    \textit{Bottom-right}: 2D stellar half-light radius; this property remains unresolved at all three levels, with measurements falling below the numerical convergence radius (see Appendix~\ref{app:resolution} for details).
    }
    \label{fig:resolution}
\end{figure*}

Fully quantifying the resolution effects on the simulation results would require a new DREAMS suite where resolution is systematically varied, which is beyond the scope of this work.  Here, we perform an initial study that varies the resolution for a single galaxy in the current suite.  While this will provide some intuition on the resolution dependence for some satellite properties, for others, one galaxy will not be enough to draw any conclusions.

Figure~\ref{fig:resolution} compares the properties of a single galaxy from the DREAMS CDM suite, simulated at three different mass and spatial resolutions. In addition to the simulation using the `Fiducial' resolution, $M_{\rm DM}=1.8\times10^6~\mdot$ and $M_{\rm \rm b}=2.8\times10^5~\mdot$, we include a new `Low Res' simulation of the same galaxy with $8\times$ more-massive particles and a `High Res' simulation with $8\times$ less-massive particles.
The Fiducial simulation has a gravitational softening of $\epsilon_{\rm grav} = 0.441$~kpc; for the Low~(High) Res simulation, this is increased~(decreased) to 0.882~(0.221)~kpc.
Figure~\ref{fig:resolution} compares the cumulative satellite mass function~(SMF)~(top left), satellite radial distribution~(top right), DM density slope~(bottom left), and stellar half-light radius~(bottom right) across these three resolution levels.

The top-left panel of Figure~\ref{fig:resolution} shows the cumulative satellite count; we display the cumulative SMF because the number of satellites present in each bin is small for this particular DREAMS host.
The Low Res simulation significantly under-predicts the number of satellites at all masses, identifying $\sim$$2\times$ fewer satellites than the Fiducial simulation between $10^7<\mstar<10^8~\mdot$.
The High Res simulation shows reasonable convergence with the Fiducial simulation; however, it predicts $\sim$1--2 more satellites than the Fiducial simulation over the same mass range.
While resolution can affect the amount of tidal stripping that a satellite experiences as it orbits the MW, these differences likely arise from shifts in the stellar mass of a given satellite. 
\cite{2018Pillepicha} showed that the stellar mass of galaxies in TNG simulations increases with resolution, and the differences we find here are consistent with their results.

The top-right panel of Figure~\ref{fig:resolution} compares the satellite radial distribution across the different resolutions.  The plot is similar to the right panel of Figure~\ref{fig:mass_func}, except that we choose to display the 3D radial distribution within 300~kpc to increase the number of satellites and clarify any trends.  (The results are similar for the 2D radial distribution, except with larger uncertainty because of the worse statistics.)
While there is some variation between the simulations, there is no clear systematic trend with resolution, at least for this specific galaxy.  

The bottom-left panel of Figure~\ref{fig:resolution} investigates how the DM density slope at 1.5~kpc, $\alphaslope$, depends on total halo mass---as previously shown in Figure~\ref{fig:slope}~(right).
Resolution has a clear impact on this property. 
The Low Res simulation finds much shallower slopes than both the  Fiducial and High Res simulations.
This is expected, as the gravitational softening for the Low Res simulation is $\epsilon_{\rm grav} = 0.882$~kpc, making $2.8\epsilon_{\rm grav}=2.5$~kpc, well outside 1.5~kpc where the slope is measured.
However, the scatter plot shows no strong systematic deviation between the Fiducal and High Res simulations for satellites with $\mtot>10^{9.5}~\mdot$.
For $10^9>\mtot>10^{9.5}~\mdot$, the satellites in the High Res simulation are systematically cuspier. 
Thus, we find that $\alphaslope$ is converged for satellites with $\mtot>10^{9.5}~\mdot$ in this galaxy.

Finally, the bottom right panel shows the 2D half-light radius, $R_\mathrm{half}$, previously shown in Figure~\ref{fig:size}.
This property is affected by two primary limitations, as discussed in Section~\ref{sec:stellar}.
First, galaxy sizes cannot be reliably measured below a convergence radius.
This limit is often taken as $2.8~\times~\epsilon_{\rm grav}=1.2$~kpc or $0.05~\times~l=1.8$~kpc~\citep{2020Ludlow}.
As noted in Section~\ref{sec:stellar}, the mean $R_\mathrm{half}$ for the satellites in the emulated DREAMS Varied dataset is all below this limit at all $\mstar$, indicating that the half-light radii are unresolved. Second, the simulations are subject to spurious numerical heating due to the use of unequal DM and stellar particles, leading to mass segregation and energy transfer to the star particles~\citep{2019Ludlow-equi}.
This numerical heating artificially inflates the stellar component, an effect more pronounced in lower-mass galaxies.
This effect likely causes the negative slope in mean $R_\mathrm{half}$ when $\mstar<10^{8.5}~\mdot$.

Comparing the Low Res, Fiducial, and High Res simulations, we confirm that the stellar half-light radii in DREAMS are not converged at the fiducial resolution.
The scatter plot shows that as resolution is increased, satellites become both more massive and smaller, moving down and to the left in the plot.
This shows the compounding effect of the increasing resolution with stellar mass discussed earlier~\citep{2018Pillepicha} and the numerical heating that affects the satellite sizes~\citep{2020Ludlow, 2019Ludlow-equi}.

\bibliography{citations}
\bibliographystyle{aasjournal}

\label{lastpage}
\end{document}